\documentclass[twocolumn]{svjour3}          
\smartqed  
\usepackage{graphicx}
\usepackage{natbib}
\usepackage{amssymb,amsmath,amsfonts,graphics,setspace,sidecap,float}
\usepackage[margin=1in]{geometry}

\usepackage{color}
\definecolor{darkgreen}{rgb}{0,0.6,0}
\definecolor{orange}{rgb}{.9,0.4,0}

\newcommand{\D}{\displaystyle}
\newcommand{\ignore}[1]{}
\renewcommand{\d}{{\rm d}}

\begin{document}

\title{Networks that learn the precise timing of event sequences}

\titlerunning{Precise timing of event sequences}

\author{Alan Veliz-Cuba \and Harel Z. Shouval \and Kre\v{s}imir Josi\'{c}* \and
 Zachary P. Kilpatrick*}
 
 \institute{A. Veliz-Cuba \at 
 Department of Mathematics, University of Houston,
              Houston TX 77204 USA \\ \email{alanavc@math.uh.edu}
\and
H.Z. Shouval \at
Department of Neurobiology and Anatomy, University of Texas Medical School, Houston TX 77030 USA \\ \email{Harel.Shouval@uth.tmc.edu}
\and
K. Josi\'{c} \at
Department of Mathematics and Department of Biology, University of Houston, Houston TX 77204 USA \\ \email{josic@math.uh.edu}
\and
Z.P. Kilpatrick \at
Department of Mathematics, University of Houston, Houston TX 77204 USA \\  \email{zpkilpat@math.uh.edu} \\[2ex] *equal contribution}


\maketitle

\begin{abstract}

Neuronal circuits can learn and replay firing patterns evoked by sequences of sensory stimuli. After training, a brief cue can trigger a spatiotemporal pattern of neural activity similar to that evoked by a learned stimulus sequence. Network models show that such sequence learning can occur through the shaping of feedforward excitatory connectivity via long term plasticity. Previous models describe how event order can be learned, but they typically do not explain how precise timing can be recalled. We propose a mechanism for learning both the order and precise timing of event sequences. In our recurrent network model, long term plasticity leads to the learning of the sequence, while short term facilitation enables temporally precise replay of events. Learned synaptic weights between populations determine the time necessary for one population to activate another. Long term plasticity adjusts these weights so that the trained event times are matched during playback. While we chose short term facilitation as a time-tracking process, we also demonstrate that other mechanisms, such as spike rate adaptation, can fulfill this role. We also analyze the impact of trial-to-trial variability, showing how observational errors as well as neuronal noise result in variability in learned event times. The dynamics of the playback process determine how stochasticity is inherited in learned sequence timings. Future experiments that characterize such variability can therefore shed light on the neural mechanisms of sequence learning.


\end{abstract}

\section{Introduction}

Networks of the brain are capable of precisely learning and replaying sequences,
accurately representing the timing and order of the constituent events \citep{Conway2001,Buhusi2005}. Recordings in awake monkeys and rats reveal neural mechanisms that underlie such sequence representation. After a training period consisting of the repeated presentation of a cue followed by a fixed sequence of stimuli, the cue alone can trigger a pattern of neural activity correlated with the activity pattern evoked by the stimulus sequence \citep{Eagleman2012,Xu2012}. Importantly, the temporal patterns of the stimulus-driven and cue-evoked activity are closely matched \citep{Shuler2006, gavornik2014}.

Sequence learning and replay has been identified in a number of different brain areas. Recent electrophysiological recordings have located patterned neural activity in V1 \citep{Xu2012,gavornik2014} and V4 \citep{Eagleman2012}, corresponding to learned visual sequences. Experiments on motor sequence learning found the underlying activity was coordinated by a combination of prefrontal, associative, and motor cortical areas \citep{jenkins1994,sakai1998}. Training networks of the brain to replay motor sequences is important since it allows quick motor skill execution, faster than deliberate muscle control allows \citep{hikosaka2002}. In addition, learning visual sequences can aid in experience-based prediction, so animals can react quickly to an unexpected chain of events \citep{meyer2011,kok2012}. Learning serial order is also an essential component of language and speech production in humans \citep{burgess1999}. In a similar way, music perception and production requires that humans learn to recognize and generate auditory-motor sequences \citep{zatorre2007}. In total, sequence learning plays a large role in the daily cognitive tasks of a wide variety of animals.

Various neural mechanisms have been proposed for learning the duration of a single event \citep{Buonomano2000,rao2001,Durstewitz2003,Reutimann2004,karmarkar2007,Gavornik2009}, as well as the order of events in a sequence \citep{Amari1972,Kleinfeld1986,Wang1990,Abbott1996,Jun2007,Fiete2010,Brea2013}.
However, mechanisms for learning the precise timing of multiple events in a sequence remain largely unexplored. The activity of single neurons evolves on the timescale of tens of milliseconds. It is therefore likely that sequences on the timescale of seconds are represented in the activity of populations of cells. Recurrent network architecture could determine activity patterns that arise in the absence of input, but how this architecture can be reshaped by training to support precisely timed sequence replay is not understood.

Long term potentiation (LTP) and long term depression (LTD) are fundamental neural mechanisms that change the weight of connections between neurons \citep{kandel2001}.  Learning in a wide variety of species, 
neuron types, and parts of the nervous system has been shown to occur through LTP and LTD \citep{Alberini2009,takeuchi2014,nabavi2014}. It is therefore natural to ask whether LTP and LTD can play a role in the learning of sequence timing \citep{karmarkar2007,ivry2008,bueti2014}, in addition to their proposed role in learning sequence order \citep{Abbott1996,Fiete2010}.

We introduce a neural network model capable of learning the timing of  events in a sequence. 
The connectivity and dynamics in the network are shaped by two mechanisms:
long term plasticity and short term facilitation. Long term
synaptic plasticity allows the network to encode sequence and timing information in the synaptic weights, while slowly evolving short term facilitation can mark time during event playback.
These ideas are quite general, and we show that they do not depend on the particulars of the time-tracking mechanisms we implemented.  The impact of stimulus variability and neural noise is largely determined by the trajectory of the time-tracking process. Thus, we predict that errors in event sequence recall may be indicative of the mechanism that encodes them.

\section{Material and Methods}

\subsection{Population rate model with short term facilitation}
\label{sec:model}

Pyramidal cells in cortex  form  highly connected clusters~\citep{song2005,perin2011}, which can correspond to neurons with similar stimulus tuning \citep{ko2011}. We therefore considered a rate model describing the activity of $N$ excitatory populations (clusters) $u_j$ ($j = 1, \ldots, N$), and a single inhibitory population $v$. The excitatory and inhibitory populations were coupled via long range connections. Each population $j$ received an external input $I_j(t)$.
Our model took the form:
\begin{align}\label{eqn:main}
\begin{split}
\tau \frac{\d u_j}{\d t} & =-u_j + \varphi\left(I_j(t)+ I_{j,syn} - \theta  \right),\\
\tau_f \frac{\d p_j}{\d t} & =1-p_j+(p_{max}-1)u_j,\\
\tau \frac{\d v}{\d t} & =-v + \varphi\left(\sum_{k=1}^N Z_{k}u_k - \theta_v\right),
\end{split}
\end{align}
where synaptic inputs to the $j$th population are given by
\begin{align*}
I_{j,syn} = w_{jj}u_j+\sum_{k\neq j}^N w_{jk}p_ku_k -Lv.
\end{align*}
A complete description of the model functions and parameters is given in Table 1. Population firing rates ranged between a small positive value (the background firing rate), and a maximal value
(the rate of a driven population), normalized to be 0 and 1, respectively. The baseline weight of the connection from population $k$ to $j$ was denoted by $w_{jk}$. Connections within a population were denoted $w_{jj}$, and these were not subject to short-term facilitation. Furthermore, the global inhibitory population, modeled by $v,$ was typically active for very short epochs, so the effects of short term plasticity were not considered. These assumptions did not change our results, but made the analysis more transparent. 

Synapses between populations were subject to short term facilitation, and the facilitated connection had ``effective synaptic strength" $w_{jk}p_k$~\citep{tsodyks1998}. Without loss of generality we assumed that short term facilitation varied between 1 and 2 so that the effective synaptic strength varied from $w_{jk}$ to $2w_{jk}$. Note that rescaling the maximal level of short term facilitation will simply rescale the relationship we will derive between the baseline synaptic weight and activation time of single populations.
We  assumed $\tau_f\gg \tau$, in keeping with the observation that synaptic facilitation dynamics are much slower than changes in firing rates~\citep{markram1998}. 

In the absence of external stimulus or input from other populations, the dynamics of each population is described by $\tau \frac{\d {u}_j}{\d t}=-u_j+\varphi(w_{jj}u_j-\theta)$, and the stationary firing rate is given by the solutions of $\varphi(w_{jj}u_j-\theta)=u_j$. We typically took $\varphi$ to be a Heaviside step function. Thus, if $w_{jj}>\theta$, there are two equilibrium firing rates: $u_j=0$ (inactive population) and $u_j=1$ (active population). This assumption simplified the
analysis, however we show show in Section~\ref{sec:altfr} that they are not essential. 

\begin{table*}
\fbox{\parbox{\textwidth}{\caption{Variables and parameters with their default values.  The default values were used in all simulations, unless otherwise noted. References indicate work where similar parameter values were used or estimated: $^1$\cite{Graupner2012}, $^2$\cite{Xu2012}, $^3$\cite{gavornik2014}, $^4$\cite{Markram1996}, $^5$\cite{Lundstrom2015}; $^6$\cite{hausser1997}. }
\label{table1}
\begin{tabular}{l|l}
\hline 
\multicolumn{2}{c}{Variables} \\
\hline
symbol & description  \\ \hline
$I_j$ & external stimulus for excitatory population $j$\\
$u_j$ & non-dimensional firing rate of excitatory population $j$ (maximum $u_j=1$) \\
$v$ & non-dimensional firing rate of global inhibitory population (maximum $v=1$) \\
$p_j$ & level of facilitation of synapses from population $j$ (baseline $p_j=1$) \\
$w_{jk}, w$ & strength of excitation from population $k$ to excitatory population $j$\\
$T_j, T$ & duration of stimulus\\
\hline 
\multicolumn{2}{c}{Time parameters (default values in parenthesis)} \\
\hline
symbol & description \\ \hline
$\tau$ & timescale of neuronal firing (10ms $^6$) \\
$\tau_f$ & timescale of short term facilitation (1s $^4$)\\
$\tau_w$ & timescale of learning rule (150s $^1$)\\
$\tau_a$ & time scale of adaptation (400ms $^5$)\\
$\tau_s$ & time scale of synaptic inputs from other populations (50ms $^6$)\\
$T_{cue}$ & duration of stimulus to trigger replay (50ms $^{2,3}$)\\
$D$ & delay in presynaptic firing affecting connections between populations (30ms $^1$)\\
$D'$ & delay in presynaptic firing affecting connections within populations (20ms $^1$)\\
\hline 
\multicolumn{2}{c}{Other parameters (default values in parenthesis)} \\
\hline
symbol & description \\ \hline
$\varphi$ & firing rate response function (Heaviside step function)\\
$\theta$ & threshold for activation of excitatory population (0.5) \\
$\theta_v$ & threshold for activation of inhibitory population (0.5)\\
$p_{max}$ & maximum level of short term facilitation (2) \\
$Z_{k}$ & strength of excitation from population $k$ to inhibitory population (0.3)\\
$L$ & weight of global inhibition (0.6) \\
$b$ & strength of adaptation (1)\\
$M$ & learning rule threshold (1)\\
$w_{max}$ & maximum synaptic weight between populations (0.4852)\\
$w'_{max}$ & maximum synaptic weight within populations (4.1312)\\
$w_{min}$ & minimum synaptic weight within populations (1.3488)\\
$\gamma_d$ & strength of LTD between populations (150 $^1$)\\
$\gamma_p$ & strength of LTP between populations (3614.5 $^1$)\\
$\gamma'_d$ & strength of LTD within populations (7500)\\
$\gamma'_p$ & strength of LTP within populations (267.86 $^1$)\\
\end{tabular}
}}
\end{table*}

\subsection{Rate-based long term plasticity}
\label{sec:training}

Connectivity between the populations in the network was subject to long term potentiation (LTP) and long term depression (LTD).  Following experimental evidence~\citep{Bliss1973,Dudek1992,Markram1996,sjostrom2001}, connections were modulated using a rule based on pre- and post-synaptic activity with  `soft' bounds \citep{Gerstner2002}.  

We made three main assumptions about the long term evolution of synaptic weight, $w = w_{\text{pre} \rightarrow \text{post}}$: (a) If the presynaptic population activity was low ($u_{pre} \approx 0$), the change in synaptic weight was negligible ($\dot{w}(t) = 0$); (b) If the presynaptic population was highly active ($u_{pre}\approx1$) and the postsynaptic population responded weakly ($u_{post}\approx0$), then the synaptic weight decayed toward zero ($\dot{w}(t) \propto -w$); and (c) If both populations had a high level of activity ($u_{post} \approx 1$ and $u_{pre} \approx 1$), then synaptic weight increased towards an upper bound ($\dot{w}(t) \propto w_{max} - w$). 

Similar assumptions have been used in previous rate-based models of LTP/LTD \citep{malsburg1973,bcm1982,oja1982,miller1994}, and it has been shown that calcium-based \citep{Graupner2012} and spike-time dependent \citep{clopath2010,gjorgjieva2011} plasticity rules can be reduced to such rate-based rules \citep{Pfister2006}. Furthermore, the fact that pre-synaptic activity is necessary to initiate either LTP or LTD is supported by experimental observations that plasticity depends on calcium influx through
NMDA receptors \citep{malenka2004}. A simple differential equation that implements these assumptions is
\begin{align}
\label{eqn:dw}
\tau_w \frac{\d w}{\d t}=&-\gamma_d \, w \, u_{pre}(t-D)(M-u_{post}(t)) \\ & +\gamma_p \, (w_{max}-w) \, u_{pre}(t-D) \, u_{post}(t), \nonumber
\end{align}
where $\tau_w$ is the time scale, $\gamma_d$ ($\gamma_p$) represents the strength of LTD (LTP), $D$ is a delay accounting for the time it takes for the presynaptic firing rate to trigger plasticity processes, and $M$ is a parameter that determines the threshold and magnitude of LTD. We note that we initially model only the molecular processes that detect correlations in firing rates, and
thus set $\tau_w = 150$s.  We will later extend this model to account for the longer timescales of synaptic weight changes (Section \ref{sec:longtimescale}).

Eq.~\eqref{eqn:dw} describes a Hebbian rate-based plasticity rule with soft bounds involving only linear and quadratic dependences of the pre- and post-synaptic rates \citep{Gerstner2002}. Temporal asymmetry that accounts for the causal link between pre- and post-synaptic activity is incorporated with a small delay in the dependence of pre-synaptic activity $u_{pre}(t-D)$ \citep{gutig2003}. This learning rule is a firing-rate version of the calcium-based plasticity model proposed by \cite{Graupner2012}.

\subsection{Encoding timing of event sequences}

Our training protocol was based on several recent experiments that explored cortical learning in response to sequences of visual stimuli \citep{Xu2012,Eagleman2012,gavornik2014}. During a training trial, an external stimulus $I_j(t)$ activated one population at a time. Each individual stimulus could have a different duration (Fig.~\ref{fig1}B). We stimulated $n$ populations, and enumerated them by order of stimulation; that is, population 1 was stimulated first, then population 2, and so on. This numbering  is arbitrary, and the initial recurrent connections have no relation to this order. We denote the duration of input $j$ by $T_j$.  All inputs stop at $T_{tot}=T_1+T_2+\ldots +T_n$. A sequence was presented $m$ times. 

Repeated  training of the network  described by Eq.~(\ref{eqn:main}) with a fixed sequence drove the synaptic weights $w_{ij}$ to equilibrium values. We assumed that during sequence presentation, the amplitude of external stimuli $I_j(t)$ was sufficiently strong to dominate the dynamics of the population rates, $u_j$. Then, the activity of the populations during training evolved according to:
\begin{align}
\tau \frac{\d u_j}{\d t} &= -u_j + \varphi\left(I_j(t)- \theta\right), \hspace{5mm} j=1,\ldots,n, \nonumber \\[0.5ex]
\tau_w \frac{\d w_{jk}}{\d t} &=-\gamma_dw_{jk}u_{k}(t-D)(M-u_{j}(t)) \label{eqn:training} \\ &+\gamma_p(w_{max}-w_{jk})u_{k}(t-D)u_{j}(t),  \hspace{1mm} j\neq k.  \nonumber
\end{align}
Thus, the timing of population activations mimicked the timing of the input sequence, \emph{i.e.} the training stimulus.

\subsubsection{Synaptic weights for consecutive activations}

For simplicity, we begin by describing the case of two populations, $N =2$,  and we consider the threshold that determines the level of LTD equal to 1, $M=1$, so that LTD is absent when the postsynaptic population is active ($u_{post}=1$). Suppose that $I_1(t)=I_S$ on $[0,T_1]$  and $I_2(t) = - I_H$, and  $I_2(t)=I_S$ on $[T_1,T_1+T_2]$ and $I_1(t) = -I_H$, where $I_S$ and $I_H$ are large enough so that Eq.~(\ref{eqn:training}) is valid. The positive inputs with weight $I_S$ model feedforward excitation to cells tuned to the cue from upstream visual processing regions in thalamus. Negative inputs with weight $-I_H$ model strong effective feedforward inhibition to cortical cells that are not tuned to the present cue \citep{wang2007,haider2013}. Representing the effect of feedforward inhibition as static inputs simplified the model, and did not affect our results. 

We assumed that $T_i>D$, $T_i \gg \tau$, and $\tau_w \gg \tau$, so that the stimulus was longer than the plasticity delay, and plasticity slower than the firing rate dynamics. Separation of timescales in Eq.~(\ref{eqn:training}) implies that  $u_j \approx \varphi ( I_j (t) - \theta)$, so the firing rate of populations 1 and 2 is approximated by $u_1(t)\approx1$ on $[0,T_1]$ and zero elsewhere, and $u_2(t)\approx1$ on $[T_1,T_1+T_2]$ and zero elsewhere (Fig.~\ref{fig2}A). Hence, during a training trial on a time interval $[0,T_{tot}]$, we obtain from Eq.~(\ref{eqn:training}) the following piecewise equation for the synaptic weight, $w_{21}$, in terms of the duration  of the first stimulus, $T_1$, 
\begin{align}
\frac{\d w_{21}}{\d t} = \left\{ \begin{array}{cl} 0, & t \not\in [D,T_1+D], \\
 \D -\frac{\gamma_d}{\tau_w} w_{21}, & t \in [D,T_1], \\
 \D \frac{\gamma_p(w_{max}- w_{21})}{\tau_w}, & t \in [T_1,T_1+D]. \end{array} \right.  \label{w2piece}
\end{align}
Note that we assumed that $u_1(t)=0$ for $t<0$.

Eq.~(\ref{w2piece}) allows the network to encode $T_1$ using the weight $w_{21}$. Namely, solving Eq.~(\ref{w2piece}) we obtain
\begin{align*}
w_{21}(T_{tot})=&w_{21}(0)e^{-T_1\gamma_d/\tau_w}e^{-(\gamma_p-\gamma_d)D/\tau_w} \\ &+ (1-e^{-D\gamma_p/\tau_w})w_{max},
\end{align*}
which relates the synaptic weight at the end of a presentation, $w_{21}(T_{tot})$, to  the synaptic weight at the beginning of the presentation, $w_{21}(0)$ (Fig.~\ref{fig2}A). Thus, there is a recursive relation that relates the weight $w_{21}$ at the end of the $i+1$st stimulus to the weight at the end of the $i$th stimulus:
\begin{align}\label{eqn:wm}
w_{21}^{i+1}=& w_{21}^i e^{-T_1\gamma_d/\tau_w}e^{-(\gamma_p-\gamma_d)D/\tau_w} \\ &+  (1-e^{-D\gamma_p/\tau_w})w_{max}. \nonumber
\end{align}
As long as $\gamma_p>\gamma_d$, the sequence $(w_{21}^i)$ converges to
\begin{equation}\label{eqn:wlimit}
 w_{21}^{\infty}:=\frac{(1-e^{-D\gamma_p/\tau_w})w_{max}}{\D 1-e^{-T_1\gamma_d/\tau_w}e^{-(\gamma_p-\gamma_d)D/\tau_w}},
\end{equation}
as seen in Fig.~\ref{fig2}B. An equivalent expression also holds in the case of an arbitrary number of populations.

The relative distance to the fixed point $w_{21}^{\infty}$ is computed by noting that (for $T_1$ fixed)
\[
| w_{21}^{i+1}- w_{21}^{\infty}|/| w_{21}^{i}- w_{21}^{\infty}|=e^{-T_1\gamma_d/\tau_w}e^{-(\gamma_p-\gamma_d)D/\tau_w},
\]
from which we  calculate
\[
| w_{21}^{i}- w_{21}^{\infty}|\propto \left(e^{-T_1\gamma_d/\tau_w}e^{-(\gamma_p-\gamma_d)D/\tau_w}\right)^i.
\]

Thus, the sequence converges exponentially with the number of training trials, $i$. The relative distance to the fixed point is proportional to $e^{-m T_1\gamma_d/\tau_w}$, so the convergence is faster for larger values of $T_1$, as shown in Fig.~\ref{fig2}C.

\subsubsection{Synaptic weights of populations that are not co-activated}
To compute the dynamics of $w_{12}$, we note that during a training trial on the time interval $[0,T_{tot}]$, the following piecewise equation governs the change in synaptic weight,
\begin{align}\notag
\frac{\d {w}_{12}}{\d t} = \left\{ \begin{array}{cl} 
0, & t \not\in [T_1+D, T_1+T_2+D], \\ 
\D - \frac{\gamma_d}{\tau_w} w_{12}, & t \in [T_1+D, T_1+T_2+D], 
\end{array} \right.
\end{align}
which can be solved explicitly to find
\begin{equation}
w_{12}(T_{tot})=w_{12}(0)e^{-T_2\gamma_d /\tau_w}.
\end{equation}

We can therefore write a recursive equation for the weight after the $i+1$st stimulus in terms of the weight after the $i$th stimulus
\[w^{i+1}_{12}=w^i_{12} e^{-T_2\gamma_d /\tau_w} ,\]
which converges to $w^{\infty}_{12}=0.$
Thus, $w^\infty_{jk}=0$ for all pairs of populations $(j, k)$ for which population $j$ was not activated immediately after population $k$ during training. 
In total, sequential activation of the populations leads to the strengthening only of the weights $w_{j+1,j},$ 
while other weights are weakened.

\subsection{Reactivation of trained networks}
\label{sec:replay}

To examine how a sequence of event timings could be encoded by our network, the first neural population in the sequence was activated with a short cue. Typically, this cue was of the form $I_1(t) = 1$ for $t \in [0,T_{cue}]$, $I_1(t)=0$ for $t \in [T_{cue}, \infty)$, and $I_j(t) = 0$ for $j\neq 1$ (Fig.~\ref{fig1}C). During replay, aside from the initial cue, the activity in the network was generated through recurrent connectivity.

We describe the case of two populations where the first population is cued, and remains active due to self-excitation ($u_1(t) \approx 1$), Fig.~\ref{fig3}A. Since $u_2(0)=0$ and $\varphi$ is the Heaviside step function, the equations governing the dynamics of the second population are
\begin{align}\notag
\tau \frac{\d {u}_2}{\d t} &=-u_2+\varphi(w_{21}p_1-\theta), \\[0.5ex]
p_1(t) &=p_{max}+(1-p_{max})e^{-t/\tau_f}.
\end{align}
Thus, for population 2 to become active, $w_{21}p_1(t)$ must have reached $\theta$ (Fig.~\ref{fig3}A). The time $T$ between when $u_1$ becomes active and $u_2$ becomes active (``replay time'') could be controlled by the synaptic weight $w_{21}$. Fig.~\ref{fig3}B shows the effect of changing the synaptic weight: For very small values of the baseline weight $w_{21}$, the effective synaptic strength $w_{21}p_1(t)$ never reaches $\theta$ and activation does not occur. Increasing the baseline weight $w_{21}$ causes more rapid activation of the second population, and for very large weights $w_{21}$ the activation is immediate. The  weight required for a presynaptic population to activate a postsynaptic population after $T$ units of time is given in closed form by 
\begin{equation}\label{eqn:W_T}
\mathcal{W}(T):=\frac{\theta}{p_{max}+(1-p_{max})e^{-T/\tau_f}}.
\end{equation}
Similarly, the inverse of this function,
\begin{equation}\label{eqn:T_W}
\mathcal{T}(w):=\tau_f \ln\left(\frac{p_{max}-1}{p_{max}-\theta/w }\right),
\end{equation}
gives the activation time as a function of the synaptic weight (Fig.~\ref{fig3}C). Note that Eq.~(\ref{eqn:T_W}) is valid for $\theta/p_{max}<w<\theta$. If $w\leq \theta/p_{max}$, then activation of the next population does not occur. If $w\geq \theta$,  activation is immediate.

To ensure the first population  becomes inactive when population 2 becomes active, we assumed that global inhibition overcame the self excitation in the first population, $w_{11}+w_{12}p_2(T)-L-\theta<0$. Also, for the second population to remain active, we needed the self excitation plus the input received from population 1 to be stronger than the global inhibition; namely, $w_{22}+w_{21}p_1(T)-L-\theta=w_{22}-L>0$.  
These two inequalities are satisfied whenever $w_{12}$ is small enough and  $L<w_{jj}<L+\theta$.

\subsection{Matching training parameters to reactivation parameters}

To guarantee that long term plasticity leads to a proper encoding of event times, it is necessary that the learned weight, $w_{21}^\infty$ given by Eq.~(\ref{eqn:wlimit}), matches the desired weight $\mathcal{W}(T)$ given by Eq.~(\ref{eqn:W_T}). This can be achieved by equating the right hand sides of Eq.~(\ref{eqn:wlimit}) and Eq.~(\ref{eqn:W_T}), so that
\begin{align}
 & \frac{(1-e^{-D\gamma_p/\tau_w})w_{max}}{1-e^{-T\gamma_d/\tau_w}e^{-(\gamma_p-\gamma_d)D/\tau_w}} = \notag \\   & \qquad \qquad =\frac{\theta/p_{max}}{1+e^{-T/\tau_f} (1-p_{max})/p_{max}}.   \label{weightmatch} 
\end{align}
 Eq.~(\ref{weightmatch}) can be satisfied for all values of $T$ by choosing parameters that satisfy
\begin{align}\label{eqn:parfit}
\begin{split}
 \tau_w/\gamma_d & =\tau_f, \\
 (1-e^{-D\gamma_p/\tau_w})w_{max} & =\theta/p_{max}, \\
 e^{-(\gamma_p-\gamma_d)D/\tau_w} & =(p_{max}-1)/p_{max}.
\end{split}
\end{align}
Since there are fewer equations than model parameters, there is a multi-dimensional manifold of parameters for which  Eq.~(\ref{weightmatch}) holds for all $T$. For instance, for fixed short-term facilitation parameters $\theta$, $p_{max}$, and $\tau_f$ and restricting specific plasticity parameters $\tau_w$ and $D$, the appropriate $\gamma_d$, $\gamma_p$, and $w_{max}$ can be determined using Eq.~(\ref{eqn:parfit}). This is 
how we determined the parameters in Figs.~\ref{fig4} and~\ref{fig5}. 

The first relationship in Eq.~(\ref{eqn:parfit}) states $\tau_w/ \gamma_d = \tau_f$, relating the timescale of short term facilitation to the timescale of long term plasticity through the depression amplitude parameter $\gamma_d$. It is important to note that this does not mean that the timescales of the two processes need to match. As stated in Table 1, following experimental data~\citep{Markram1996,Alberini2009,Graupner2012,nabavi2014},  we chose $\tau_f=1$s and $\tau_w=150$s for our simulations. This implies that $\gamma_d = 150$, for training to yield the correct weights.

As we demonstrate in Supplementary Fig. 1, perturbing parameters of the long term plasticity process away from the optimal relationships determined by Eq. (12) does alter the learned time. The relative size of errors depends on the parameters we perturb, and we find the model is most sensitive to perturbation of $w_{max}$. Perturbations of other parameters such as $\gamma_p$ and $\gamma_d$ lead to to errors roughly equal in to the magnitude of the parameter perturbation  (e.g., a 5\% perturbation of $\gamma_d$ leads to a 5\% change in $T_{replay}$).

For more detailed models, the analog of Eq.~(\ref{eqn:parfit}) is more cumbersome or impossible to obtain explicitly. Specifically, when we incorporated noise into our models in the Section \ref{sec:effnoise} and considered spike rate adaptation in the Section \ref{sec:slowadapt}, we had to use an alternative approach. We found it was always possible to use numerical means to approximate parameter sets that allowed a correspondence between the trained and desired weight for all possible event times. A simple way of finding such parameters was to use the method of least squares: We selected a range of stimulus durations, e.g. $[.1s,3s]$, and sampled timings from it, e.g. $S=\{.1s,.2s,.3s,$ $\ldots, 3s\}$. For each $T\in S$ we computed the learned weight, $w_{learned}(T,\text{pars})$, where ``pars'' denotes the list of parameters to be determined. Then, we computed the replay time, $T_{replay}(w_{learned}(T,\text{pars}))$. We defined the ``cost'' function
\[
J(\text{pars})=\sum_{T\in S}\left(T_{replay}(w_{learned}(T,\text{pars}))-T\right)^2,
\]
and the desired parameters were given by
\begin{align}\label{eqn:numerical_parfit}
\text{pars}_{best}:=argmin_{\text{pars}}\{J(\text{pars})\}.
\end{align}
This  approach was successful for different models and training protocols, and allowed us to find 
a working set of parameters for models that included noise or different slow processes for tracking time.

Eq.~(\ref{eqn:numerical_parfit}) can also be interpreted as a learning rule for the \emph{network parameters}. Starting with arbitrary network parameter values, any update mechanism that  decreases the cost function $J$ will result in a network that can accurately replay learned sequence times.

\subsection{Training and replay simulations}
\label{sec:training_replay}

To test our model, we trained the network with a sequence of four events.  Each event in the sequence corresponded to the activation of a single neuronal population (Fig.~\ref{fig4}A). Since each population was inactivated (received strong negative input) when the subsequent populations became active, we also assumed that an additional, final population inactivated the population responding to the last event (additional population not shown in the figure). Input during training was strong enough so that activation of the different populations was only determined by the external stimulus overriding global inhibition and recurrent excitation. 

During reactivation, the recurrent connections were assumed fixed. This assumption can be relaxed if we assume that  LTP/LTD are not immediate, but occur on long timescales, as in the Section~\ref{sec:longtimescale}.

We used $m=10$ training trials, the default  parameter values in Table~\ref{table1}, and estimated $\gamma_d$, $\gamma_p$, and $w_{max}$ using Eq.~(\ref{eqn:parfit}). After the training trials were finished, we cued the first population in the sequence, using $I_1(t)=1$ for $t \in [0,T_{cue}]$ and $I_j(t)=0$ otherwise.  We also started with this set of weights, and retrained the network with a novel sequence of stimuli.

\section{Results}

\begin{figure*}[t]
\centering
\includegraphics[width=14cm]{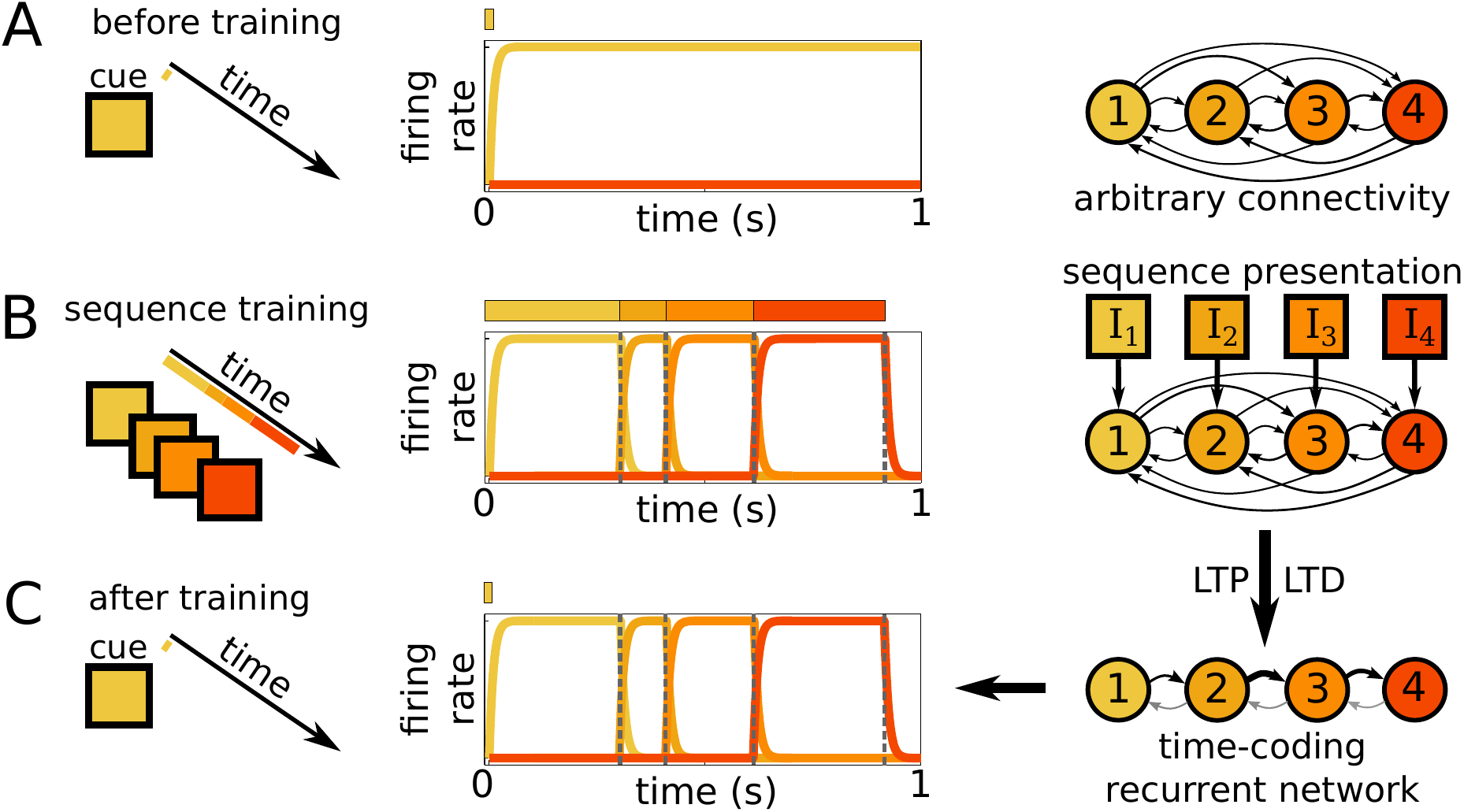}
\caption{The precise timing of training sequences is learned via long term plasticity.  Each stimulus in the sequence is represented by a different color.
{\bf A}. Before training, network connectivity is random, and a cue does not trigger a sequential pattern of activity.
{\bf B}. During training, a sequence of events is presented repeatedly. 
Each event activates a corresponding neural population for some amount of time, which is fixed across presentations.
Long term plasticity reshapes network architecture to encode the duration and order of these activations. 
{\bf C}. After sufficient training, a cue triggers the pattern of activity evoked during the training period. Learned synaptic connectivity along with short term facilitation steer activity along the path carved by  the training sequence (arrow width and contrast correspond to synaptic weight).}
\label{fig1}
\end{figure*}

\begin{figure}[t]
\centering
\vspace{0cm}
\includegraphics[width=6.4cm]{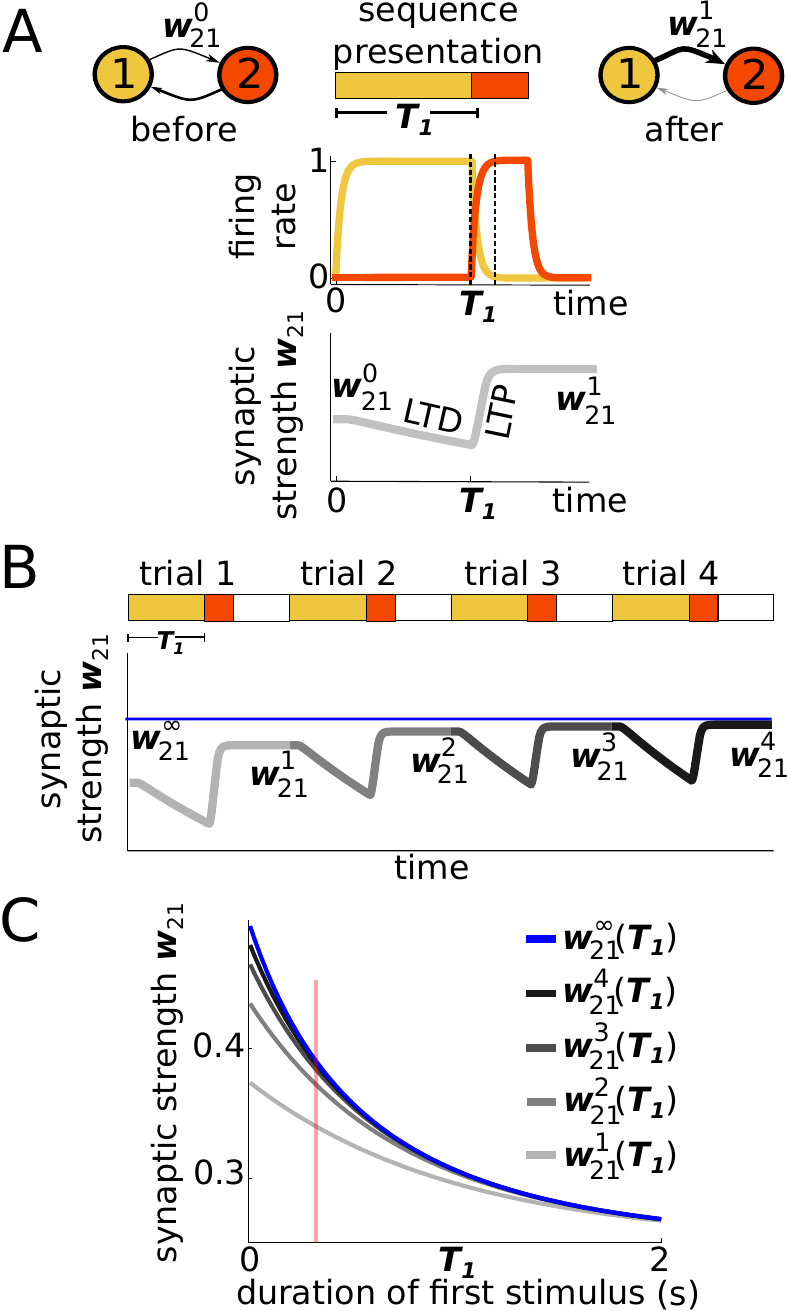}
\caption{Encoding timing in synaptic weights.
{\bf A}. Synaptic connections evolve during training. When a presynaptic population (1) is active and a postsynaptic population (2) is inactive, LTD reduces the synaptic weight $w_{21}$.  When the populations (1 and 2) are co-active (overlap window between dashed lines), LTP increases $w_{21}$. Shortly after, global inhibition inactivates the presynaptic population (1), so long term plasticity ceases (see Materials and Methods). As in the text, $w^i_{21}$ denotes  the weight at the end of the $i$-th trial. Arrow width and contrast correspond to synaptic weights.
{\bf B}. After several training trials, the synaptic weight $w_{21}^i$ converges to a fixed point, $w^{\infty}_{21}$, whose amplitude depends on the activation time of the presynaptic population.
{\bf C}. Starting from the same initial value, $w^0_{21}(T_1) = 0.25,$ the weight  $w^i_{21}(T_1)$ converges to different values, $w^{\infty}_{21}(T_1)$, depending on the the training time, $T_1$. Pink bar at $T_1$=300ms corresponds to the value used in {\bf A} and {\bf B}. 
}
\label{fig2}

\end{figure}

\subsection{Training}

We explore sequence learning in a network model of neural populations, where each population is activated by a distinct stimulus or event.  The initial connectivity between the populations is random (Fig. \ref{fig1}A).  To make our analysis more transparent, we initially consider a deterministic firing rate model, Eq.~(\ref{eqn:main}). Each individual neural population is bistable, having both a low activity state and a high activity state that is maintained through recurrent excitation.  Our results also hold for more biologically plausible firing rate response functions and are robust to noise (Sections \ref{sec:effnoise} and \ref{sec:altfr}). 

To train the network, we stimulated populations in a fixed order, similar to the training paradigm used by \cite{Xu2012,Eagleman2012,gavornik2014}. The duration of each event in the training sequence was arbitrary (Fig. \ref{fig1}B), and each stimulus in the sequence drove a single neural population. Synaptic connections between populations were plastic.  To keep the model tractable, population activity was assumed to immediately impact the weight of synaptic connections. Our results also extend to a model with synaptic weights changing on longer timescales (Section \ref{sec:longtimescale}).

Changes in the network's synaptic weights depended on the firing rates of the pre- and post-synaptic populations \citep{Bliss1973,bcm1982,Dudek1992,Markram1996,sjostrom2001}. When a presynaptic population was active, either: (a) synapses were potentiated (LTP) if the postsynaptic population was subsequently active or (b) syna\-pses were depressed (LTD) if the postsynaptic population was not activated soon after  (Materials and Methods). Such rate-based plasticity rules can be derived from spike time dependent plasticity rules \citep{Kempter1999, Pfister2006, clopath2010}. 

To demonstrate how the timing of events can be encoded in the network architecture, we start with two populations (Fig. \ref{fig2}). During training, population 1 was stimulated for $T_1$ seconds followed by stimulation of population 2 (Fig. \ref{fig2}A). The stimulus was strong enough to dominate the dynamics of the population responses (Materials and Methods). While the first stimulus was present, population 1 was active and LTD dominated, decreasing the synaptic weight, $w_{21}$, from population 1 to population 2.
After $T_1$ seconds, the first stimulus ended, and the second population was activated. However, population 1 did not become inactive instantaneously, and for some time both population 1 and 2 were active.  During this overlap window, LTP dominated leading to an increase in synaptic weight $w_{21}$. Shortly after population 1 became inactive, changes in the weight $w_{21}$ ceased, as plasticity only occurs when the presynaptic population is active. The initial and final synaptic weights ($w^0_{21}$ and $w^1_{21}$, respectively) can be computed in closed form (Materials and Methods). Repeated presentations of the training sequence leads to exponential convergence of the synaptic weights, $w^i_{21}$ (weight after $i$th training trial), to a fixed value  (Fig. \ref{fig2}B).  On the other hand, the synaptic weight $w_{12}$ is weakened during each trial because the presynaptic population 2 is always active after the postsynaptic population 1 (Materials and Methods).   In the case of $N$ populations, each weight $w_{k+1,k}$ will converge to a nonzero value associated with $T_k$, whereas all other weights will become negligible during replay. Thus, the network's structure eventually encodes the order of the sequence.

The duration of activation in population 1, $T_1,$ determines the equilibrium value of the synaptic weight from population 1 to population 2, $w^{\infty}_{21}$ (Materials and Methods). For larger values of $T_1$, LTD lasts longer, weakening $w_{21}$ (Fig. \ref{fig2}C). Hence, weaker synapses are associated with longer event times. Reciprocally, weaker synapses  lead to longer activation times during replay (Section \ref{sec:slowprocess}).

As the stimulus duration, $T_1$, determines the asymptotic  synaptic weight, $w^{\infty}_{21}(T_1)$, there is a mapping $T_1 \rightarrow w^{\infty}_{21}(T_1)$ from  stimulus times to the resulting weights. Event timing is thus encoded in the asymptotic values of the synaptic weights.

\begin{figure}[t]
\centering
\vspace{0cm}
\includegraphics[width=6.7cm]{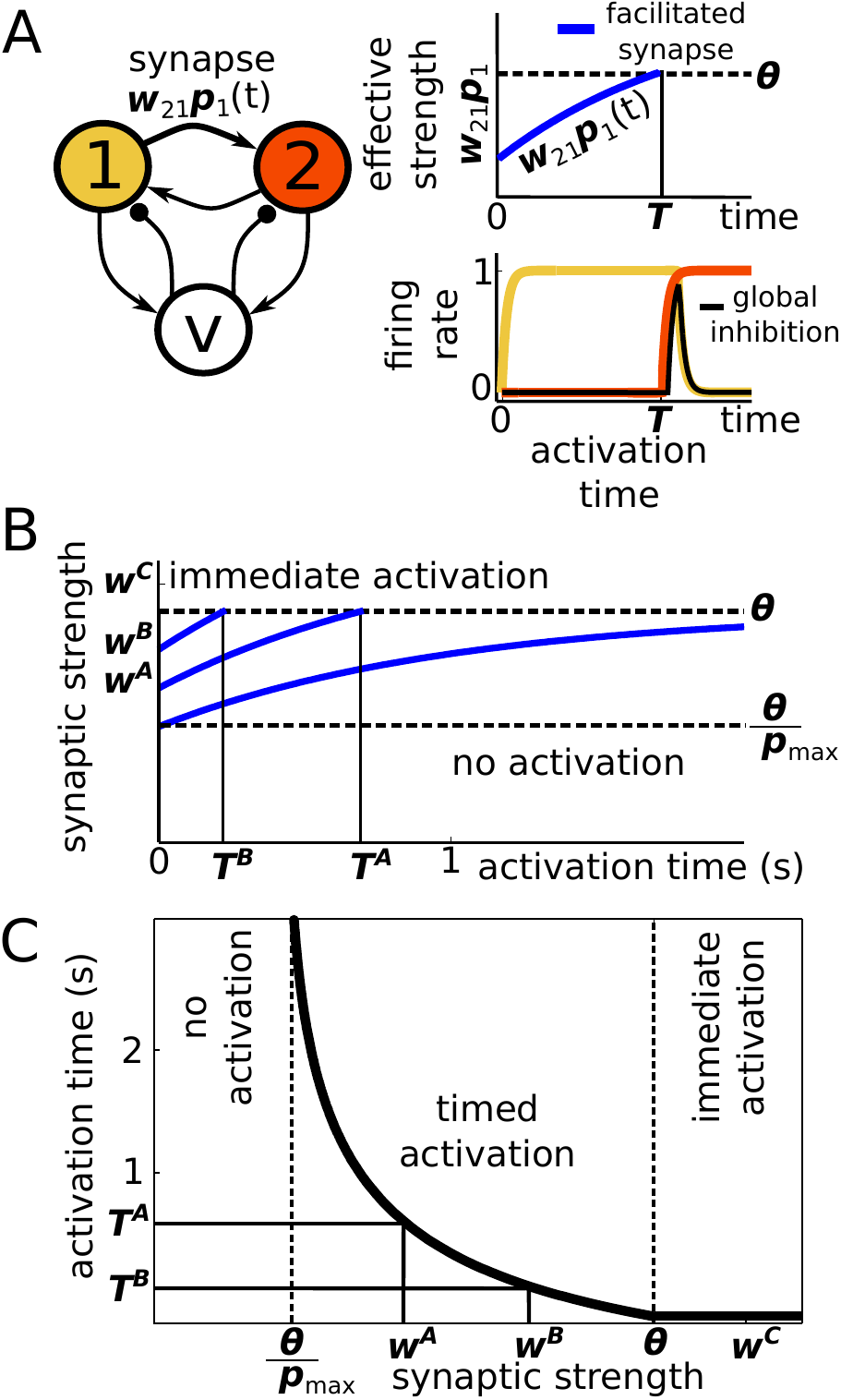}
\caption{Replay timing. 
{\bf A}. Once population 1 is activated, the weight of the connection to  population 2 slowly increases, eventually becoming strong enough to activate this next population in the sequence. 
{\bf B}. Activation time $T:=T_1$ decreases monotonically with the weight of the connection between the populations, $w:=w_{21}$. For weak connections ($w^A=0.33$) the synapses must be strongly facilitated to reach the threshold. If the weight is larger ($w^B=0.42$) threshold is reached more quickly. For weights above the threshold ($w^C=0.58>\theta$), population 2 is activated immediately. Activation will not occur when the synaptic connection is smaller than $\theta/p_{max}$.
{\bf C}. Activation time $T_1$ plotted against the initial synaptic weight, $w_{21}$. For intermediate values of $w_{21}$, the relationship is given by Eq.~(\ref{eqn:T_W}). Here $w^A, w^B, w^C$ and $T^A,T^B$ are the same as in panel~{\bf B}.}
\label{fig3}
\end{figure}

\subsection{A slow process allows precise temporal replay}
\label{sec:slowprocess}

We next describe how the trained network replays sequences. The presence of a slow process, which we assumed here to be short term facilitation, is 
critical. This slow process tracks time by {\em ramping} up until reaching a pre-determined threshold. An event's duration corresponds to the amount of time it takes the slow variable to reach this threshold.   Such ramping models have previously been proposed as mechanisms for time-keeping\citep{Buonomano2000,Durstewitz2003,Reutimann2004,karmarkar2007,Gavornik2009}. Without such a slow process, cued activity would result in a sequence replayed in the proper order, but information about event timing would be lost.

For simplicity we focus on two populations,  whe\-re activity of the first population represents a timed event (Fig. \ref{fig3}). To simplify the analysis, we also assumed  that synaptic weights are fixed during replay. This assumption is not essential (Section \ref{sec:longtimescale}). After population 1 is activated with a brief cue, it remains active due to recurrent excitation (Materials and Methods). Meanwhile, short term facilitation leads to an increase in the effective synaptic strength from population 1 to population 2.  Population 2 becomes active when the input from population 1 crosses an activation threshold (Fig. \ref{fig3}A). When both populations are simultaneously active, a sufficient amount of global inhibition is recruited to shut off the first population, which receives only weak input from population 2.  The second population then remains active, as the strong excitatory input from the first population and recurrent excitation exceed the global inhibition.

The weight of the connection from population 1 to population 2 determines how long it takes to extinguish the activity in the first population (Fig. \ref{fig3}B).  This synaptic weight therefore encodes the time of this first and only event. We demonstrate how this principle extends to multiple event sequences in the Section \ref{sec:timecoding}. The time until the activation of the second population decreases as the initial synaptic weight increases, since a shorter time is needed for facilitation to drive the input from population 1 to the activation threshold  (Fig. \ref{fig3}C). Note that when the baseline synaptic weight is too weak, synaptic facilitation saturates before the effective weight reaches the activation threshold, and  the subsequent population is never activated.  When the baseline synaptic weight is above the activation threshold, the subsequent population is activated instantaneously.

Therefore, long term plasticity allows for encoding an event time in the weight of the connections between populations in the network, while short term facilitation is crucial for replaying the events with the correct timing. The time of activation during cued replay will match the timing in the training sequence as long as training drives the synaptic weights to the value that corresponds to the appropriate event time (Fig. \ref{fig2}C):

\begin{align}
\mathcal{W}(T):=\frac{\theta}{p_{max}+(1-p_{max})e^{-T/\tau_f}}.  \label{res_wt}
\end{align}
Here $p_{max}$ and $\tau_f$ are the strength and timescale of facilitation (Materials and Methods). An exact match can be obtained by tuning parameters of the long term plasticity process so the learned weight matches Eq.~(\ref{res_wt}). There is a wide range of parameters for which the match occurs (Materials and Methods). We next show that the timing and order of sequences containing multiple events can be learned in a similar way.

\begin{figure}[t]
\centering
\includegraphics[width=7.3cm]{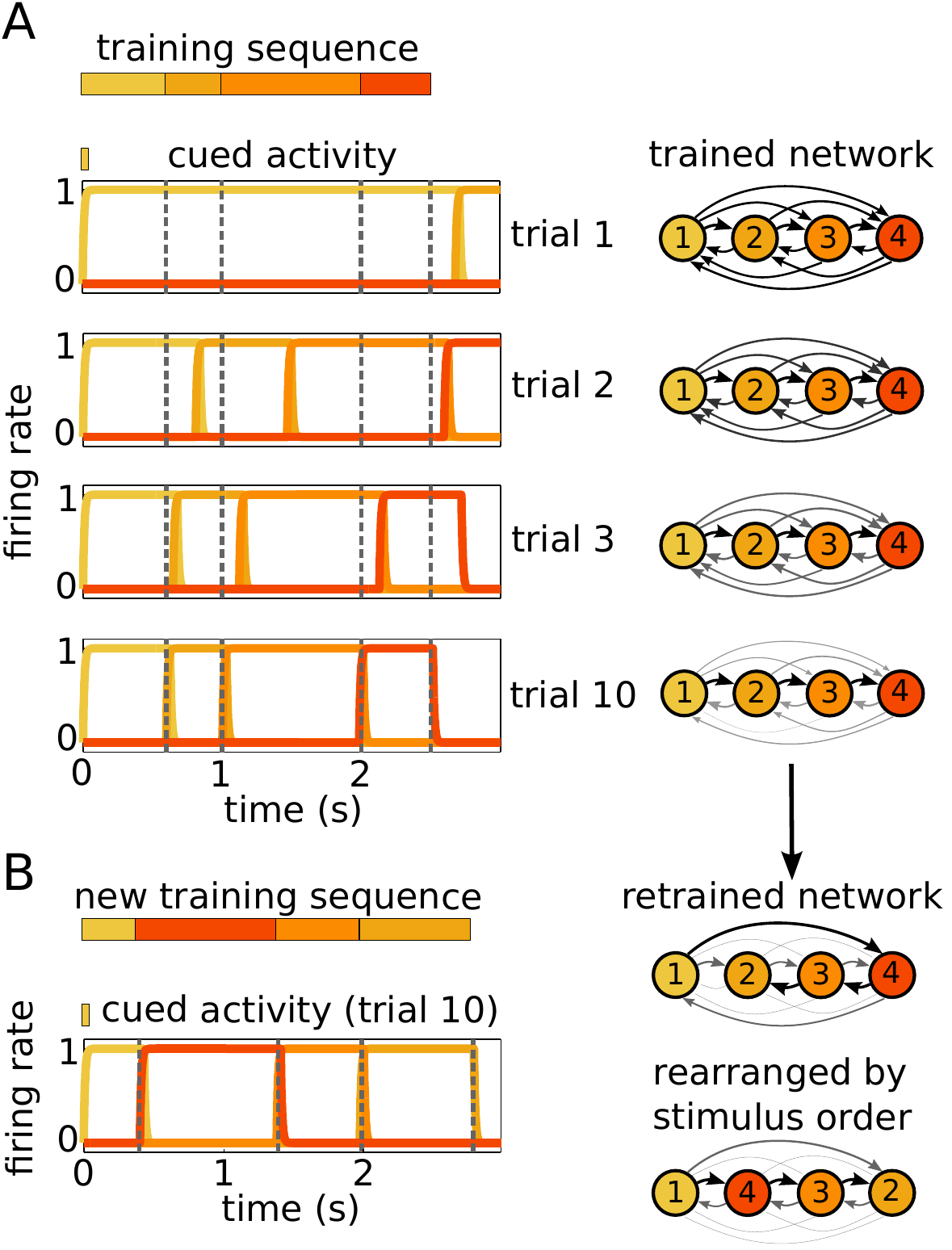}
\caption{Learning and replaying event sequences.
{\bf A}. As the number of training trials increases, a cue results in an activation pattern that approaches that evoked by the training sequence. Network architecture is reshaped to encode the precise duration and order of events in the sequence, with stronger feedforward connections corresponding to shorter events (thicker and darker arrows correspond to larger synaptic weights). All weights are learned independently and training most strongly affects the weights $w_{i\rightarrow i+1}$. The event times were $T_1=0.6$, $T_2=0.4$, $T_3=1$, and $T_4=0.5$ seconds, with indices denoting the population stimulated. The initial weights were $w_{ij}^0=0.025$ for $i\neq j$ and $w_{ii}=1$.
{\bf B}. The network in the last row of {\bf A} was retrained with the sequence  $T_1=0.4$, $T_4=1$, $T_3=0.6$, and $T_2=0.8$, presented in the order 1-4-3-2.  After 10 training trials, the cued network replays the new training sequence.}
\label{fig4}
\end{figure}

\subsection{Repeated presentation of the same sequence produces a time-coding feedforward network}
\label{sec:timecoding}

To demonstrate that the mechanism we discussed extends easily to arbitrary sequences, we consider a concrete sequence of four stimuli. We set the parameters of the model  so that the training parameters match the reactivation parameters (Eqs. \ref{weightmatch}-\ref{eqn:numerical_parfit} in Materials and Methods). 

We trained the network using the event sequence 1-2-3-4 (Fig.~\ref{fig4}A),  repeatedly  stimulating the corresponding populations  in succession. The duration of each population activation was fixed across trials. After each training trial, we cued the network by stimulating the first population for a short period of time to trigger  replay (Materials and Methods). Thus, after population 1 is activated  the subsequent activity is  governed by the network's architecture. As our theory suggests, the cue-evoked network activity pattern converged with training to the stimulus-driven activity pattern  (Fig. \ref{fig4}A). 

We further tested whether the same network can be retrained to encode a sequence with a different order of activation (1-4-3-2) with different event times. Fig. \ref{fig4}B shows that after training, the network encodes and replays  the new training sequence. Thus, the network architecture can be shaped by long term plasticity  to encode an arbitrary sequence of event times, and a brief cue evokes the replay of the learned sequence. 

\section{Extensions}

\subsection{Effects of noise}
\label{sec:effnoise}

\begin{figure}[t]
\vspace{0cm}
\centering
\includegraphics[width=7.7cm]{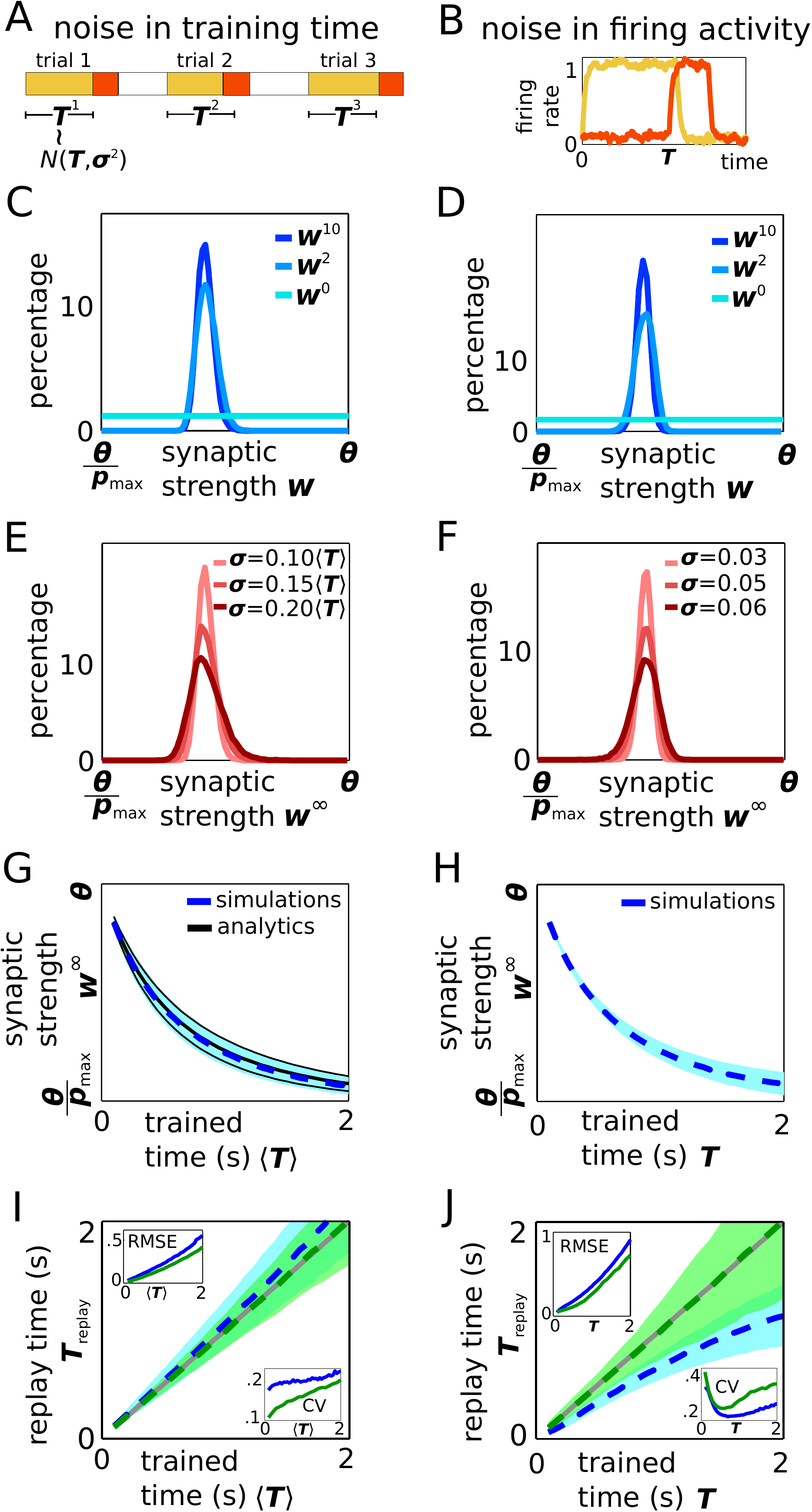}
\caption{Effect of noise on learning and replay (caption continued on next page).
}
\label{fig5}
\end{figure}
\setcounter{figure}{4}

\begin{figure}[t]
\caption{Effect of noise on learning and replay. The effects of adding normally distributed noise to {\bf A} the observed training time $T:=T_1$ (duration of first stimulus) and {\bf B} neural activity $u_j$. Starting from a uniform distribution ($p(w^0_{21})$) and considering either a noise level of {\bf C} $\sigma=0.1\langle T \rangle$ in the training time or {\bf D} $\sigma=0.03$ in neural activity, the probability density function $p(w^i_{21})$ converges to the steady state distribution $p(w^{\infty}_{21})$ (this is nearly identical to $w^{10}_{21}$). Increasing noise in the  training times ({\bf E}) or  neural activity ({\bf F})  widens the steady state distribution $p(w^{\infty}_{21})$. The mean (dashed lines) and standard deviation (shaded region) of $w^{\infty}_{21}(T)$ are pictured in panel {\bf G} when  noise with standard deviation $\sigma=0.20 \langle T \rangle$ is added to the training times and {\bf H} when  noise with standard deviation $\sigma=0.05$ is added to neural activity. As in Fig. \ref{fig2}C, the mean learned weight $E \left[ w^{\infty}_{21} \right]$ decreases with $\langle T \rangle$. The mean of the replay time (dashed blue line) and its standard deviation (shaded cyan region) are plotted against $\langle T \rangle$ for the case of {\bf I} noise in training times and {\bf J} noise in neural populations. As the mean training time $\left\langle T \right\rangle$ increases, so does the effect of noise on replay time, $T_{\text{replay}}$ (grey lines show the diagonal line $T_{replay}= T$). For a suitable choice of ``corrected-for-noise'' parameters, the effect of noise on the mean replay time can be removed (dashed green line and green shaded region). Insets in {\bf I} and {\bf J} show the root-mean-square error in replay time as a function of training time for the  parameters obtained from the deterministic case (blue) and corrected-for-noise parameters (green).
}
\end{figure}

To examine the impact of the many sources of variability in the nervous system~\citep{faisal2008}, we explored how noise impacts the training and recall of event sequences in our model. We examined the effects of stochasticity in event times as well as noise in the network activity and the impact such variability has on the training and recall of event sequences. For simplicity, we focus on the case of two populations, but our results extend to sequences of arbitrary length. 

To examine the impact of variability in stimulus durations, we sampled $T:=T_1$ from a normal distribution (Fig.~\ref{fig5}A) with mean $\left\langle T\right\rangle=0.5$ and variance $\sigma^2=(c_v\left\langle T\right\rangle)^2$, where $c_v=0.1$ is the coefficient of variation. Randomness in the observed event time may be due to variability in the external world, temporal limitations on sensation \citep{butts2007}, or other observational errors \citep{ma2006}. We selected $w^0:=w^0_{21}$ from a uniform distribution on $[\theta/p_{max},\theta]$. Fig.~\ref{fig5}C shows the evolution of the probability density function of $w^m$ as $m$ increases (using 20,000 initial $w^0$'s). The synaptic weight after the $i$th training, $w_{21}^i,$ is described by a probability density function that converges in the limit of many training trials. The peak (mode) of this distribution is the most likely value of the learned synaptic weight after repeated presentation of the sequence (Fig.~ \ref{fig5}C). The variance of  the learned synaptic weight, $w^{\infty}_{21},$ increases monotonically with the variance in the training time, $\sigma^2$ (Fig. \ref{fig5}E). We explored the effects of different levels of noise, $\sigma=0.1\langle T \rangle, 0.15\langle T \rangle, 0.2\langle T \rangle$ (20,000 initial conditions for each), and estimated the probability density function of $w^{\infty}$ numerically (Fig.~\ref{fig5}E). To see the effect of noise for different mean training durations, we estimated the mean and standard deviation of $w^{\infty}$ for $\left\langle T \right\rangle$ ranging from $0.1$s to $2$s (step size of 0.05s) using $\sigma=0.2\langle T \rangle$  (5,000 initial conditions for each) (Fig.~\ref{fig5}G). For a distribution of training times with mean $\langle T_1 \rangle$ we obtain a unimodal probability density for the weights, $p(w_{21}).$ As in the noise-free case, the mode of this weight distribution decreases with $\langle T_1 \rangle$. Note that the parameters given by Eq.~(\ref{eqn:parfit}), which guarantee that training and replay time coincide in the deterministic case, may not be the same as the parameters needed when noise is present. We numerically estimated these parameters using Eq.~(\ref{eqn:numerical_parfit}) so the mean training time and mean replay time coincided.

To determine how noise affects activation timing during sequence replay, we compared the mean event time with the mean replayed time. Since the  network parameters used here are those obtained from the noise-free case, we expect that replay times are biased. Indeed, Fig.~\ref{fig5}I shows that activity during replay is slightly longer  on average than the corresponding training event. Also, the variance in activation during replay increases with the mean duration of the trained event. We can search numerically and find a family of parameters for which the mean activation time during replay and training coincide (Fig.~\ref{fig5}I). Error and the coefficient of variation (CV) in replay time increases with the duration of the trained time (Fig.~\ref{fig5}I, shaded region and inset).

We also estimated the mean learned synaptic weight and its variance analytically:  
Since the synaptic weight, $w^i_{21}$,  evolves according to the rule 
\[w^{i+1}_{21}=w^i_{21} A(T^i) +  C,\]
where $A(T):=e^{-T\gamma_d/\tau_w}e^{-(\gamma_p-\gamma_d)D/\tau_w}$ and $C:=(1-e^{-D\gamma_p/\tau_w})w_{max}$, we obtain 
\[
E[w^{i+1}_{21}]=E[w^i_{21} A(T^i)] +  C.
\]

Since $w^i_{21}$ only depends on $T^j$ for $j<i$, it follows that $w^i_{21}$ and $A(T^i)$ are independent random variables; then,
\[
E[w^{i+1}_{21}]=E[w^i_{21}] E[A(T^i)] +  C,
\]
and by taking the limit $i\rightarrow \infty$ and solving for $\lim_{i\rightarrow \infty}{E[w^i_{21}]}$ we find
\[
\mu_w:=\lim_{i\rightarrow \infty}{E[w^i_{21}]}=\frac{C}{1-\mu_A},
\]
where $\mu_A:=E[A(T^i)]$. Squaring $w^{i+1}_{21}=w^i_{21} A(T^i) +  C$ gives
\[(w^{i+1})^2=(w^i)^2 A^2(T^i) +  C^2 + 2w^i_{21} A(T^i)C,\]
and then it similarly follows that
\begin{align*}
\sigma_w^2:=&\lim_{i\rightarrow \infty}{\left(E[(w^i_{21})^2]-E[w^i_{21}]^2\right)} \\ =&
\frac{C^2 \sigma_A^2}{(1-\mu_A)^2(1-\sigma_A^2-\mu_A^2)},
\end{align*}
where $\sigma_A^2:=E[A(T^i)^2]-E[A(T^i)]^2$.  
To quantify the average error, we computed numerically the mean and the standard deviation of the replay time (Fig.~\ref{fig5}I). The root-mean-square error (RMSE) was computed by 
\[\text{RMSE}(\left\langle T\right\rangle):=\sqrt{E[(\left\langle T\right\rangle-T_{replay})^2]},\]
where the expected value is taken over the replay time, $T_{replay}$. The coefficient of variation (CV) was computed by
\[
CV ( \langle T \rangle ) : = \frac{\sqrt{Var(T_{replay})}}{ \langle T_{replay} \rangle}.
\]

To introduce neural noise (Fig.~\ref{fig5}B), we added white noise to the rate equations of the populations during training and replay so that

\begin{align}\notag
 {\rm d} u_j = \frac{1}{\tau}\left[ - u_j + \varphi \left( I_j(t) + I_{j,syn} - \theta \right) \right] {\rm d} t + \sigma {\rm d} \xi_j,
\end{align}
where ${\rm d} \xi_j$ is a standard white noise process with variance $\sigma^2$. The analysis of the effect of noise in population activity is similar to the analysis performed on stimulus duration noise, the only difference being that the noise level was $\sigma=0.03$ in panel Fig. \ref{fig5}D, $\sigma=0.03, 0.05, 0.06$ in
panel Fig.~\ref{fig5}F, and  $\sigma=0.05$ in panels Fig. \ref{fig5}H and \ref{fig5}J. After repeated presentation of a sequence, the distribution of the learned synaptic weights converged (Fig. \ref{fig5}D). The variance of the synaptic weight increased monotonically with the variance of the noise (Fig. \ref{fig5}F), and the mean weight decreased monotonically with the event time (Fig. \ref{fig5}H). Since we used the parameters found from the noise-free case, we expect some bias in replay time. After training, the replayed event times are shorter than the corresponding events in the training sequence (Fig. \ref{fig5}J), and the effect is much more significant than the lengthening of times due to observation noise. This systematic bias in the replayed time error is due to the saturating nature of the time-tracking process, short term facilitation \citep{markram1998}. Input to the second population remains close to threshold for longer periods of time for longer trained times, leading to more frequent noise-induced threshold crossings \citep{gardiner2004}. However, parameters for which mean event time and mean replay time coincide can be found numerically (Materials and Methods). Interestingly, the CV attains a minimum in the vicinity of the timescale of short term facilitation ($\tau_f = 1$s), suggesting the network best encodes events on the timescale of the slow process (Fig.~\ref{fig5}J, lower inset). This principle holds across a range of short term facilitation timescales $\tau_f$ (Supplementary Material, Fig. 2).


\subsection{Alternative firing rate response function}
\label{sec:altfr}

\begin{figure}[t]
\centering
\includegraphics[width=7.6cm]{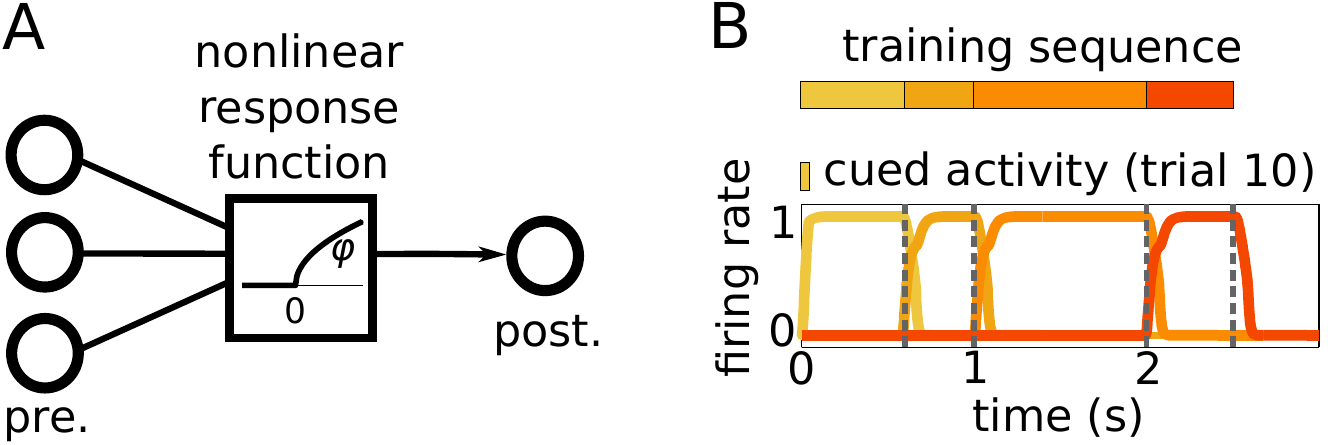}
\caption{Alternative firing rate function. The mechanism for sequence learning and replay also works for other firing rate functions (see Materials and Methods). {\bf A}. Using a nonlinear and nonsaturating response function $\varphi_\gamma(u)=\sqrt{\gamma u}$, long term plasticity still results in the coding of a training sequence as synaptic weights. {\bf B}. After training, the cued network replays the training sequence similarly to the replay seen in Fig.~\ref{fig4}A. }
\label{fig6}
\end{figure}

We next show that the mechanism for learning the precise timing of an event sequence does not depend on the particulars of the model.  In previous sections, we used a Heaviside step function as the firing rate function and chose short term facilitation as the slow, time-tracking process. However, the principles we have identified do not depend on these specific choices. More general circuit models of slowly ramping units can learn and replay timed event sequences. The elements needed to learn and replay precise time durations are a chain of slow accumulators and a learning rule which modifies weights to precisely time a threshold-crossing event for each accumulator (Supplementary Material, Fig. 3).

Precise replay relies on a threshold-crossing process which occurs as long as each population is bistable, having only low and high activity states rather than graded activity. Indeed,  detailed spiking models \citep{litwinkumar2012} and experimental recordings \citep{major2004} suggest that cell assemblies can exhibit multiple stable states. To test whether our conclusions hold with different firing rate response functions, we replace the Heaviside function with the nonsaturating function \citep{fourcaud2003} (Fig. \ref{fig6}A)
\[
\varphi_{\gamma}(x)=\begin{cases}
0 & \textrm{ if } x<0,\\
\sqrt{\gamma x} & \textrm{ if } x\geq 0,
\end{cases}
\]
where the parameter $\gamma$ determines the steepness. Note that in the absence of other population inputs,
\[
 \tau \frac{\d {u}_j}{\d t}  =-u_j + \varphi_{\gamma} \left(w_{jj}u_j- \theta\right),
\]
which has steady states determined by the equation $u=\varphi_{\gamma}\left(w_{jj} u - \theta\right)$. One of the stable steady states is $u=0$ and there is a positive  stable steady state, $u_*$, which is the largest root of the quadratic equation $u^2 - \gamma w_{jj} u + \gamma \theta = 0$.  For simplicity, we normalize  $\gamma$ and self-excitation  so that the stable states are $u=0$ and $u=1$; namely, we consider $\gamma':=\gamma/u_*^2$ and $w'_{jj}:=w_{jj}u_*$. 

Since a population is activated when its input reaches the threshold due to short term facilitation, the derivations that led to Eqs.~\ref{eqn:W_T} and \ref{eqn:T_W} are still valid for this model. However, the activation of a neuronal population ($u_j \to 1$) was delayed since $\varphi_{\gamma}$ has finite slope. This delay was negligible when  firing rate response was modeled by a Heaviside  function, and activation was instantaneous. To take this 
delay into account, we can modify  Eqs.~\ref{eqn:W_T} and \ref{eqn:T_W} to obtain
\begin{align*}
\mathcal{W}(T):=\frac{\theta}{p_{max}+(1-p_{max})e^{-(T-d)/\tau_f}},
\end{align*}
and
\begin{align*}
\mathcal{T}(w):=d+\tau_f \ln\left(\frac{p_{max}-1}{p_{max}-\theta/w} \right),
\end{align*}
where $d$ is a heuristic correction parameter to account for the time it takes for $u_j$ to approach 1.

Following the arguments that led to Eq.~(\ref{eqn:parfit}), we were able to derive constraints on parameters to ensure the correct timings are learned:
\begin{align}
\frac{\tau_w}{M\gamma_d} & = \tau_f, \nonumber \\
 \left(
1-\frac{e^{-D(\frac{(M-1)\gamma_d+\gamma_p}{\tau_w})}}{1+\frac{(M-1)\gamma_d}{\gamma_p}} 
 \right) w_{max} & = \theta/p_{max}, \label{eqn:parfit_sqrt} \\
  e^{(\gamma_d-\gamma_p)D/\tau_w}  & = \frac{p_{max}-1}{p_{max}} e^{d/\tau_f}. \nonumber
\end{align}
For simulations we used  the parameters $M=1.5$, $d=30$ms, $\gamma=3$, and estimated $\gamma_d$, $\gamma_p$, and $w_{max}$ using Eq.~(\ref{eqn:parfit_sqrt}) (we then normalized $\varphi_\gamma$ and $w_{jj}$  to make 0 and 1 the stable firing rates). As in the previous simulations, the number of presentations was $m=10$; the durations of the events were $0.6$s, $0.4$s, $1$s, and $0.5$s for events 1, 2, 3 and 4, respectively. Network architecture converges, and the replayed activity matches the order and timing of the training sequence (Fig. \ref{fig6}B). 

\subsection{An alternative slow process: spike rate adaptation}
\label{sec:slowadapt}

\begin{figure}[t]
\vspace{0cm}
\centering
\includegraphics[width=6.4cm]{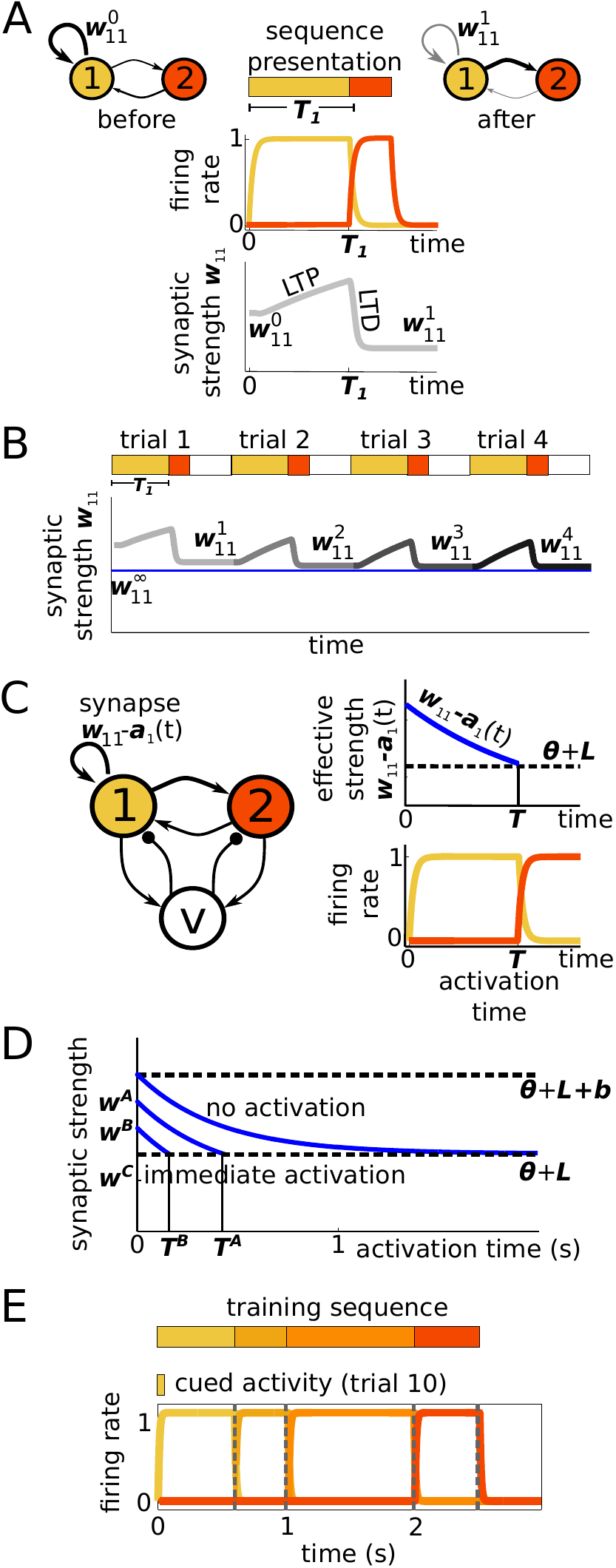}
\caption{Alternative slow process based on spike frequency adaptation (caption continued on next page).}
\label{fig7}
\end{figure}

\setcounter{figure}{6}
\begin{figure}[t]
\centering
\caption{Alternative slow process based on spike frequency adaptation. 
\textbf{A.} When a population is active, LTP increases the synaptic weight $w_{11}$. After becoming inactive, LTD decreases $w_{11}$ (see Materials and Methods). \textbf{B.} Synaptic weight $w_{11}^i$ is updated after the $i$th training trial.  After several trials, $w_{11}^i$ converges to a fixed point $w_{11}^\infty$ that depends on the activation time of the first population. \textbf{C.} Once population 1 is active, adaptation builds up until overcoming  self excitation. This will occur when the effective strength, $w_{11}-a_1(t)$ ([excitation]$-$[adaptation]), crosses below $\theta + L$ ([threshold]$+$[inhibition]). \textbf{D.} Activation time $T:=T_1$ increases with synaptic weight $w:=w_{11}$. For strong self excitation ($w^A$) adaptation takes longer to shut off the first population, so the next population in the sequence is activated later. Weaker self excitation ($w^B$) will result in quicker extinction of activity in the first population, result in the next population activating sooner. For self excitation below $\theta+L$ ([threshold] $+$ [inhibition]), the first population will inactivate immediately, resulting in immediate activation of the next population. If self excitation of the first population is greater than $\theta+L+b$ ([threshold] $+$ [inhibition] $+$ [maximum adaptation]), it will remain active indefinitely, and the subsequent population is never activated. \textbf{E.} When the parameters of the long term plasticity process and the replay process  match, the network can learn the precise timing of sequences.}
\label{fig7caption}
\end{figure}

We also examined whether spike frequency adaptation, \emph{i.e.} a slow decrease in firing rates in response to a fixed input to a neural population, can play the role of a slow, time tracking process~\citep{benda2003}, instead of short term facilitation.
In contrast to the case of short term facilitation, adaptation causes the effective input from one population to decrease over time.

In this case population activity was modeled by
\begin{align}\notag
\begin{split}
\tau \frac{\d {u}_j}{\d t} &  =-u_j+ \varphi(w_{jj}u_j+s_j -\theta-Lv-a_j),\\[0.5ex]
\tau_a \frac{\d {a}_j}{\d t} & =-a_j+b u_j,\\
\tau_s \frac{\d {s}_j}{\d t} & = -s_j+ \sum_{k\neq j}^N w_{jk}u_k,\\
 \tau \frac{\d {v}}{\d t} & =-v + \varphi\left(\sum_{k=1}^N Z_{k}u_k - \theta_v\right),
\end{split}
\end{align}
 where $a_j$ denotes the adaptation level of population $j$,  $\tau_a$ is the time scale of adaptation, and $b$ is the adaptation strength. Feedback between populations was assumed to be slower than feedback within a population; thus, the total input for population $j$ was split into self-excitation ($w_{jj}u_j$), and synaptic inputs from other populations ($s_j$) which evolved on the time scale $\tau_s$. Note that in the limit $\tau_s \to 0$, synapses are instantaneous.

For a suitable choice of parameters, global inhibition tracks activity faster than excitation between populations. Then, when a population becomes inactive due to adaptation, the level of global inhibition decreases, allowing subsequent populations to become active. This means the weight of self excitation  can encode timing. Thus, in this setup we modeled long term plasticity within a population as well. The learning rule for $w_{jj}$ was analogous to $w_{jk}$ with the additional assumption that since $w_{jj}$ represented the synaptic weight within a population,  it could not decrease below a certain value $w_{min}$. Also, the parameters for long term plasticity within a population are allowed to be different from the parameters for long term plasticity between populations.

The learning rule  was then
\begin{align*}
\tau_w\frac{\d {w}_{jj}}{\d t}= &-\gamma'_d(w_{jj}-w_{min}) u_j(t-D')(1-u_j(t)) \\ & -\gamma'_p(w_{jj}-w'_{max}) u_j(t-D')u_j(t).
\end{align*}

When the population was activated ($u_1(t) \approx 1$) for $t \in [0,T_1]$ (Fig. \ref{fig7}A), the changes in the weight $w_{11}$ were governed by the piecewise differential equation
\begin{align}\notag
\frac{\d {w_{11}}}{\d t} = \left\{ \begin{array}{cl} 0, & t \not\in [D',T_1+D'] \\ 
\D \frac{\gamma'_p}{\tau_w} ( w'_{max} - w_{11}) & t \in [D',T_1] \\ 
\D - \frac{\gamma'_d}{\tau_w} ( w_{11}- w_{min} ) & t \in [T_1, T_1+D']. \end{array} \right.
\end{align}
The following equation relates the synaptic weight at the end of a presentation, $w_{11}(T_{tot})$, to the synaptic weight at the beginning of the presentation, $w_{11}(0)$:
\begin{align*}
w_{11}(T_{tot})=&w_{11}(0) e^{-T_1\gamma'_p/\tau_w} e^{(\gamma'_p-\gamma'_d)D'/\tau_w} \\ &+
w'_{max}e^{-D'\gamma'_d/\tau_w}(1-e^{-(T_1-D')\gamma'_p/\tau_w}) \\ &+w_{min}(1-e^{-D'\gamma'_d/\tau_w}).
\end{align*}
 This recurrence relation between the weight at the $i+1$st stimulus, $w_{11}^{i+1}$, and the weight at the $i$th stimulus, $w_{11}^i$, implies that $w_{11}^i$ converges to the limit (Fig. \ref{fig7}B)
\begin{align*}
 w^\infty_{11}=&
\frac{w'_{max}e^{-D'\gamma'_d/\tau_w}(1-e^{-(T_1-D')\gamma'_p/\tau_w})}
{1-e^{-T_1\gamma'_p/\tau_w} e^{(\gamma'_p-\gamma'_d)D'/\tau_w}} \\ 
&+ \frac{w_{min}(1-e^{-D'\gamma'_d/\tau_w})}{1-e^{-T_1\gamma'_p/\tau_w} e^{(\gamma'_p-\gamma'_d)D'/\tau_w}}.
\end{align*}
Thus, for each stimulus duration a unique synaptic weight is learned. Also, as shown in previous sections, $w_{21}$ will converge to a fixed value and $w_{12}$ is weakened.  Note that in this case timing will be encoded as the weight of self-excitation, and the order will be encoded as the weights between populations.

During replay a cue activates population 1, $u_1=1$, and we obtain $\D \tau \frac{\d {u}_1}{\d t}=-u_1+\varphi(w_{11}-\theta-a-L)$, $\D \tau_a \frac{\d {a}_1}{\d t}=-a+b$. Population $1$ will become inactive ($u_1\approx 0$) when $w_{11}-a_1(t)$ decreases to $\theta+L$. Then, the next population will become active due to the decrease in global inhibition and the remaining feedback from the first population due to the slower dynamics of feedback between populations (Fig. ~\ref{fig7}C). 

The precise time of activation can be controlled by tuning the synaptic weight $w_{11}$. Furthermore, since the activation time satisfies $w_{11}-a_1(T)=\theta+L$, we have a formula that relates the synaptic weight to the activation time (Fig.~\ref{fig7}D)
\[w_{11}=\mathcal{W}(T):=\theta+L+b(1-e^{-T/\tau_a}).\]
 When self excitation is too strong, adaptation will not affect the activity of the first population and deactivation will never occur. On the other hand, when self excitation is too weak, activation is not sustained and the population will be shut off immediately. To guarantee correct time coding and decoding, $w_{11}^{\infty}(T)$ and $\mathcal{W}(T)$ had to be approximately equal for all $T$. The appropriate  parameters could not be found in closed form, so we again resorted to finding them numerically using  Eq.~(\ref{eqn:numerical_parfit}).

For simulations we used the parameters $Z_{k}= 0.6$, $L= 0.8$,   $\gamma_p=3750$, $\gamma_d=100$, $\gamma'_p=267.86$, $\gamma'_d=7500$, $w_{max}=1.5$, $w'_{max}=4.1312$,  and $w_{min}=1.3488$.  As in the previous simulations, the number of presentations was $m=10$; the duration of the events were $0.6$s, $0.4$s, $1$s, and $0.5$s for events 1, 2, 3 and 4, respectively.

This idea generalizes to any number of events and populations. Timing is encoded in the weight of the excitatory self-connections within a population, while sequence order is encoded in the weight of the connections between populations. Moreover, for a range of network parameters, the duration of the sequences during training and reactivation coincide (Materials and Methods). Presenting the event sequence used in Fig. \ref{fig4}A, the network can learn the precise timing and order of the events (Fig. \ref{fig7}E).






\subsection{Incorporating long timescale plasticity}
\label{sec:longtimescale}

Thus far, we have assumed that during sequence replay, synaptic connections remained unchanged (Section \ref{sec:timecoding}). However, if synaptic changes occur on the same timescale as the network's dynamics, and are allowed to act during replay, the network's architecture can become unstable. This problem can be solved by assuming that  synaptic weights change slowly compared to network dynamics  \citep{Alberini2009}.

\begin{figure}[t]
\centering
\includegraphics[width=7cm]{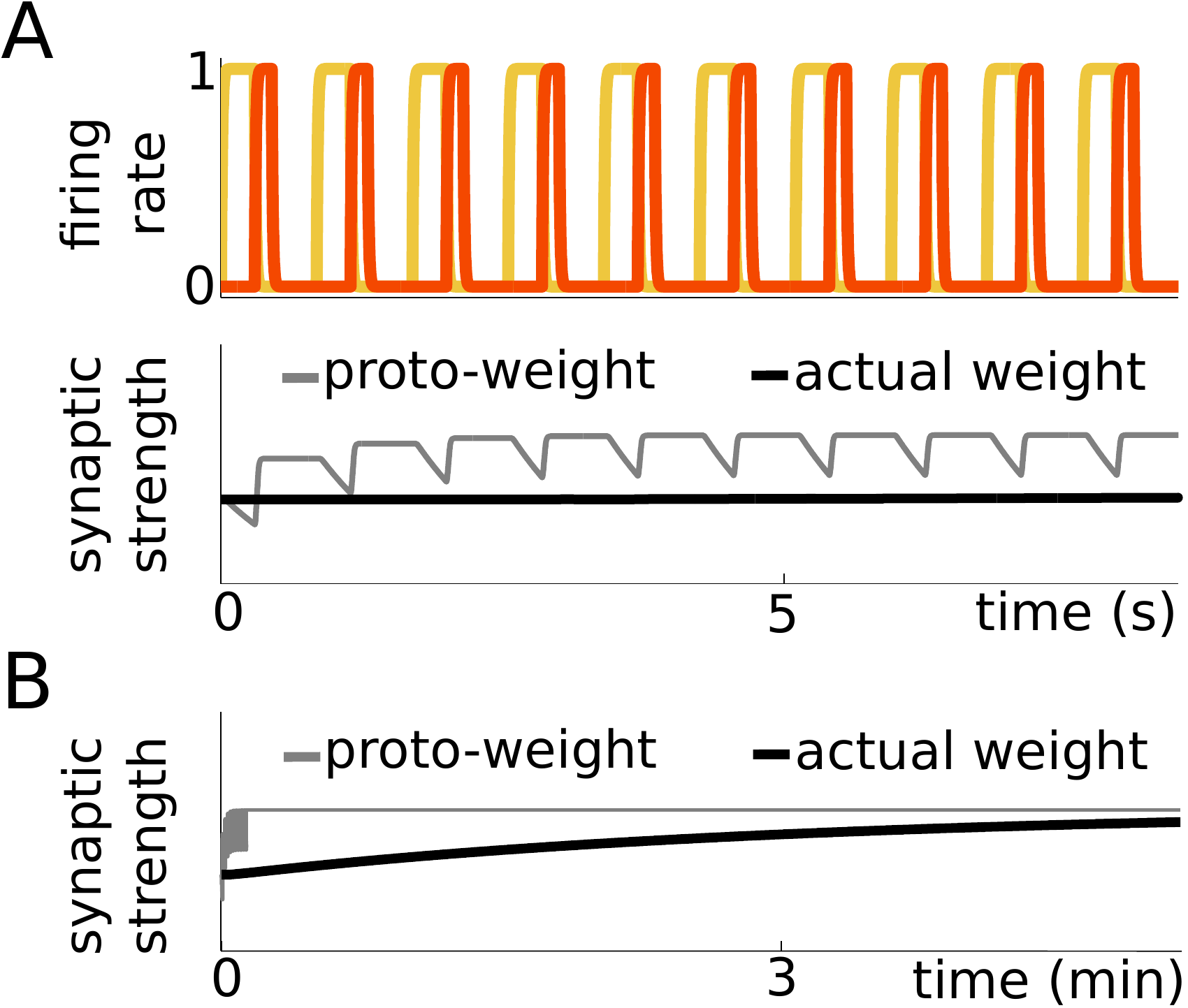}
\caption{A two-stage learning rule. {\bf A}. {\em Proto-weights} track the signaled change to synaptic connectivity brought about by firing rate covariance. {\em Actual weights} evolve on a slower time scale and hence do not track changes in the proto-weights immediately. {\bf B}. Actual synaptic weights approach proto-weight values on a longer timescale, so both eventually converge to the same value.}
\label{fig8}
\end{figure}

We therefore extended our model so that long term plasticity occurs on more realistic timescales. The impact of rate covariation on the network's synaptic weights was modeled by a two step process: (a) rate correlation detection, which occurs on the timescale of seconds and (b) translation of this information into an actual weight change, which occurs on the timescale of minutes or hours. The initial and immediate signal shaped by the firing rates of pre- and postsynaptic neural populations was modeled by intermediate variables we refer to as {\em proto-weights}~\citep{Gavornik2009} (Fig.~\ref{fig8}A).  Changes to the {\em actual synaptic weights} occur on a much longer timescale, and slowly converged to values determined by the proto-weights (Fig.~\ref{fig8}B). 



Introducing proto-weights leads to repeated reenforcement of learned activity patterns making them robust to spontaneous network activations. The model takes the form
\begin{align}\label{eqn:proto}
\begin{split}
\tau \frac{\d {u}_j}{\d t} & =-u_j + \varphi\left(I_j(t)+I_{j,syn} - \theta\right),\\
\tau_f \frac{\d {p}_j}{\d t} & =1-p_j+(p_{max}-1)u_j,\\
\tau \frac{\d {v}}{\d t} & =-v + \varphi\left(\sum_{k=1}^N Z_{k}u_k - \theta_v\right),\\
\tau_w\frac{\d {w}_{jk}}{\d t} & =-\gamma_dw_{jk}u_{k}(t-D)(M-u_{j}(t)) \\ &+\gamma_p(w_{max}-w_{jk})u_{k}(t-D)u_{j}(t),  \hspace{1mm} j\neq k,\\[0.5ex]
\tau_I\frac{\d {W}_{jk}}{\d t} & =w_{jk}(t-D_p)-W_{jk}(t),
\end{split}
\end{align}
where
\begin{align*}
I_{j,syn} = W_{jj}u_j+\sum_{k\neq j}^N W_{jk}p_ku_k -Lv,
\end{align*}
$\tau_I$ is the time scale of the actual weights, and $D_p$ is a time delay in the process of transforming changes in the proto-weights to changes in the actual weights. Since $\tau_w$ represents the timescale of changes in proto-weights, the timescale $\tau_I$ represents the timescale of actual synaptic weight changes. The process of translating a coincidence in firing rates into a weight change can be much longer than 
detecting the coincidence, and we can thus take $\tau_I$ much larger than $\tau_w$~\citep{Markram1996}.

It was still possible to analyze the model given in Eqs.~(\ref{eqn:proto}), but the  results were less transparent. During training, the proto-weights satisfied Eq.~(\ref{eqn:wm}). As the number of training trials increased,  the proto-weights converged to the limit given in  Eq.~(\ref{eqn:wlimit}). The actual weights also slowly  approached the same limit; that is, $W_{jk}$ converged to the limit given in  Eq.~(\ref{eqn:wlimit}). During replay, the proto-weights evolved according to Eqs.~\ref{eqn:proto}, but due to the delay $D_p$ and the time scale $\tau_I$, the actual weights remained unchanged and replay occurred accurately as in the Section \ref{sec:training_replay}.


Modeling long term plasticity as a two-stage process allowed us to incorporate more realistic details: We were able to assume that synaptic connections are plastic during replay, as well as during training. Replaying the sequence of neural population activations evokes long term plasticity signals through the proto-weights, and alterations to actual synaptic weights do not take place until after the epochs of neural activity. During reactivation, the actual weights already equal the values necessary to elicit the timed event sequence. Then, the weakening of the connection due to LTD will precisely equal the strengthening due to LTP, resulting in no net change in the actual connection weight. As long as the time scale of synaptic weight consolidation is much larger than the sequence timescale, long term plasticity during replay reinforces the learned network of weights that is already present.

\section{Discussion}

Sequences of sensory and motor events can be encoded in the architecture of neuronal networks. These sequences can then be replayed with correct order and timing when the first element of the sequence is presented, even in the absence of any other sensory input. Experimental evidence shows that after repeated presentations of a cued training sequence, the presentation of the cue alone triggers a temporal pattern of activity similar to that evoked by the training stimulus \citep{Eagleman2012,Xu2012,Shuler2006, gavornik2014}. Our goal here was to provide a biologically plausible mechanism that could govern the learning of precisely timed sequences.

\subsection{Learning both the precise timing and order of sequences}

We demonstrated how a complex learning task can be accomplished by combining two simple mechanisms. First, the {\em timing} of a single event can be represented by a slowly accumulating positive feedback process \citep{Buonomano2000, Durstewitz2003, Reutimann2004,SheaBrown2006,karmarkar2007, Gavornik2009,simen2011}. Second, rate dependent long term plasticity can reshape synaptic weights so that the {\em order} and precise timing of events in a sequence is encoded by the network's architecture~\citep{Amari1972,Abbott1996}.

To make the problem analytically tractable, we considered an idealized model of neural population firing rates, long term plasticity, and short term facilitation. This allowed us to obtain clear relationships between parameters of the time-tracking process (short term facilitation) and the learning process (long term plasticity).  The assumptions about model structure and parameters that were essential
for sequence learning could be explicitly described in this model.  Similar conditions were required for learning in more realistic models, which incorporated the long timescale of LTP/LTD.

A novel feature of our network model is that long term plasticity influences the length of time a neural population is active. Typical computational models of sequence learning employ networks of neurons \citep{Jun2007,Fiete2010,Brea2013} or populations \citep{Abbott1996} that are each active for equal amounts of time during replay. However, sensory and motor processes can be governed by networks whose neurons have a fixed stimulus tuning \citep{Xu2012,gavornik2014}. Therefore, a sequence of events of varying time lengths should be represented by neural populations that are each active for precisely the length of time of the corresponding event. Our model demonstrates that this can be achieved using rate-based long term plasticity.

\subsection{Experimental predictions}

The general mechanisms we described here imply a number of experimentally testable features of the neural substrates of the learning and recall of event sequences. Our analysis of the impact of noise on time encoding demonstrates a relationship between the dynamical mechanism for encoding and error statistics. If time interval estimation is accomplished through a slow process that saturates toward a threshold, then {\em the relative error} of an interval in the sequence should increase with interval length, and average interval estimate should be shortened during replay (Fig. \ref{fig5}J). This provides an innate mechanistic explanation for underestimate of time duration, contributing to existing literature that has found environmental conditions that can lead to such systematic errors \citep{morrone2005,terao2008}. When time is marked by a slow process that scales linearly in time, average duration estimates will be close to the true estimate and {\em relative error} will be invariant to trained durations (Supplementary Material, Fig. 3). We suggest a way in which average interval estimates may be shorter than the trained interval, if the trained interval is longer than the timescale of the slow process encoding it. If the slow process that reads out the stored time grows linearly or exponentially, average interval estimates may be nearly equal or longer than the true duration (Supplementary Material, Fig. 3).

The mechanics of sequence learning could be understood further by examining the development of sequence replay accuracy with the number of trainings. Errors in sequence recall will tend to be greatest after very few trainings (Fig \ref{fig4}A). Cortical recordings reveal that, indeed, the correlation between replayed activity and training sequence evoked activity increases with the number of sequence exposures \citep{Eagleman2012}. Additionally, our model suggests that errors in the replayed time of each event's beginning will build serially, if they rely upon a sequence of population activations. This means that errors in the total run time of a sequence will increase with sequence duration, as in \cite{hass2008}. However, errors in the individual estimates of each event duration will not depend on their placement in the sequence. Such errors should decrease at a similar rate across all individual events, as suggested by our analysis in Materials and Methods. This prediction could be tested experimentally by examining how subjects' individual event duration estimates depend on the event's position in a learned sequence.

Lastly, we predict that event sequences can be learned through rate correlation based synaptic plasticity acting on connection between stimulus-tuned populations. This mechanism could be probed experimentally in a number of ways. First, if neural activity underlying sequence learning were being recorded electrophysiologically \citep{Xu2012,gavornik2014}, subsequent experiments could be performed to see if electrically stimulating neurons out of sequence could disrupt learned sequence memory. This would provide evidence that plasticity mechanisms that result from neural activity are involved in the consolidation of sequence memory. Inactivating populations in the sequence could also disrupt the replay of the remainder of the sequence, supporting our network chain model of sequence learning. For example, optogenetic methods could be used to inactivate a large fraction of cells that respond to one of the events in sequence. If such a disruption were to terminate sequence replay, this would be strong evidence for the events being represented by a chain of active populations. Furthermore, long term plasticity processes could be disrupted through the local injection of translational inhibitors \citep{Alberini2009}. If this leads to a reduction of sequence memory robustness, it would constitute strong evidence for the importance of long term plasticity in the local circuit for  sequence memory formation.

\subsection{Comparison to previous models of interval timing}

Several previous theoretical studies have proposed neural mechanisms for the learning and recall of timed events. Models capable of representing serial event order have utilized individual units that are oscillators \citep{brown2000} or bistable populations \citep{grossberg1992}. Recent studies have found that continuous temporal trajectories can be learned in networks of chaotic elements by training weights to downstream neurons that constitute a linear readout \citep{buonomano2009,hennequin2014}. A complementary approach has been used to infer time by fitting a maximum likelihood model to the rates and phases of spiking neurons in hippocampal networks \citep{itskov2011}. Our approach is most similar to previous studies that utilize discrete populations or neurons to represent serial order \citep{grossberg1992,Abbott1996,Fiete2010,Brea2013}. Namely, we assume that the memory of each individual event duration is learned in parallel with the others as in \cite{Fiete2010}, in contrast to the serial building of chains demonstrated in the model of \cite{Jun2007}. {\em Reset} models of sequence replay are supported by comparisons of human behavioral data to models that mark event durations using a clock that is reset after each event \citep{mcauley2003}, suggesting errors are made locally in time, rather than accumulated event-to-event. We have extended previous work by developing a mechanism for altering the activation time of each unit in the sequence. This learning process is distinct from the approach outlined in \cite{buonomano2009,hennequin2014}, since it solely trains the recurrent architecture between populations encoding time; tuning of a downstream readout is unnecessary.

\subsection{Internal tuning of long term plasticity parameters}

There is a large set of parameters for which the network can be trained to accurately replay training sequences. 
While some parameter tuning is required, in simple cases we could find these parameters explicitly.
In all cases appropriate
parameters could be obtained  computationally using gradient descent. In biological systems plasticity processes could be shaped across many generations by evolution, or within an organism during development.
Indeed, recent experimental evidence suggests that networks are capable of internally tuning long term plasticity responses through {\em metaplasticity} \citep{abraham2008}.  For instance, NMDA receptor expression can attenuate LTP \citep{Huang1992,Philpot2001}, while metabotropic glutamate receptor activation can prime a network for future LTP \citep{Oh2006}. We note that such mechanisms would affect the timescale and features of LTP/LTD, not the synaptic weights themselves.

\subsection{Models that utilize ramping processes with different timescales}

Our proposed mechanism relies on a ramping process that evolves on the same timescale as the training sequence. Short term facilitation \citep{markram1998} as well as rate adaptation \citep{benda2003} can fulfill this role. However, other ramping processes that occur at the cellular or network level are also capable of marking time (Supplementary Material, Fig. 3). For instance, slow synaptic receptor types such as NMDA can slowly integrate sensory input \citep{wang2002}, resulting in population firing rate ramping similar to experimental observations in interval timing tasks \citep{Xu2014}. Were we to incorporate slow recurrent excitatory synapses in this way, the duration of represented events would be determined by the decay timescale of NMDA synapses. Alternatively, we could have also employed short term depression as the slow process in our model. Mutual inhibitory networks with short term depression can represent dominance time durations that depend on the network's inputs, characteristic of perceptual rivalry statistics \citep{laing2002}. This relationship between population inputs and population activity durations could be leveraged to represent event times in sequences. 

Events that occur on much shorter or longer timescales than those we explored here could be marked by processes matched to those timescales. For instance, fast events may be represented simply using synaptic receptors with rapid kinetics, such as
AMPA receptors \citep{clements1996}. AMPA receptor states evolve on the scale tens of milliseconds, which would allow representation of several fast successive events. However, we would expect a lower bound on the duration of an event represented by this mechanism, given by the neuronal membrane time constant \citep{dayan2001}. Slow events could also be represented by a long chain of sub-populations, each of which is activated for a shorter amount of time than the event. In the context of our model, this would mean each population would contain sub-populations connected as a feedforward chain \citep{goldman2009}.   Networks of cortical neurons can have different subpopulations with distinct sets of timescales, due to the variety of ion channel and synaptic receptor kinetics \citep{ulanovsky04,bernacchia2011,pozzorini2013,costa2013}. This {\em reservoir} of timescales could be utilized to learn events whose timings span several orders of magnitude.

\subsection{Learning the repeated appearance of an event}

We only considered training  sequences in which no event appeared more than once (e.g. 1-2-3-4). If events appear multiple times (e.g. 1-2-1-4), then a learned synaptic weight (e.g., $w_{21}$) would be weakened when the repeated event appears again.  This can be resolved by representing each event repetition by the activation of a different subpopulation of cells. There is evidence that this occurs in hippocampal networks responding to spatial navigation sequences on a figure eight track \citep{Griffin2007}.  Even for networks where each stimulus activates a specific population, sequences with repeated stimuli could be encoded in a deeper layer of the underlying sensory or motor system.
The same idea can be used to create networks that can store several different event sequences containing the same events  (e.g. 1-2-3-4; 2-4-3-1; 4-3-2-1). If multiple sequences begin with the same event (e.g. 1-2-3-4; 1-3-2-4), evoking the correct sequence would require partial stimulation of the sequence (e.g. 1-2 or 1-3). Networks would then be less likely to misinterpret one learned sequence for another sequence with overlapping events \citep{Abbott1996}. 

\subsection{Feedback correction in learned sequences}

We emphasize that we did not incorporate any mechanisms for correcting errors in timing during replay. However, this could easily be implemented by considering feedback control via a stimulus that activates the population that is supposed to be active, if any slippage in event timing begins to occur. This assumes there is some external signal indicating how accurately the sequence is being replayed. For instance, human performance of a piece of music relies on auditory feedback signals that are used by the cerebellum to correct motor errors \citep{zatorre2007,kraus2010}. If feedback is absent or is manipulated, performance deteriorates \citep{finney2003,pfordresher2003}. Similar principles seem to hold in the replay of visual sequences. \cite{gavornik2014} showed that portions of learned sequence are replayed more accurately when preceded by the correct initial portion of the learned sequence. We could incorporate feedback into our model by providing external input to the network at several points in time, not just the initial cue stimulus.

\subsection{Conclusions}

Overall, our results suggest that a precisely timed sequence of events can be learned by a network with long term synaptic plasticity. Sequence playback can be accomplished by a ramping process whose timescale is similar to the event timescales. Trial-to-trial variability in training and neural activity will be inherited by the sequence representation in a way that depends on the learning process and the playback process. Therefore, errors in sequence representation provide a window into the neural processes that represent them. Future experimental studies of sequence recall that statistically characterize these errors will help to shed light on the neural mechanisms responsible for sequence learning.


\begin{acknowledgements}
We thank Jeffrey Gavornik helpful comments.  Funding was provided by NSF-DMS-1311755 (Z.P.K.); NSF/NIGMS-R01GM104974 (A.V-C. and K.J.); and NSF-DMS-1122094 (A.V-C. and K.J.).
\end{acknowledgements}

\bibliographystyle{spbasic}
\bibliography{sequence_refs_v0}
\pagebreak

\end{document}